\documentclass[a4paper,onecolumn,aps,pra,nofootinbib,superscriptaddress]{revtex4-2}
\usepackage[utf8]{inputenc}
\usepackage{amsmath} 
\usepackage{mathtools}
\usepackage{amsfonts} 
\usepackage{amssymb}
\usepackage{graphics}
\usepackage{graphicx}
\usepackage{float}
\usepackage{color}
\usepackage{braket}
\usepackage{tikz}
\usepackage{pgfplots}
\usepackage{bbold}
\usepackage{hyperref}
\usepackage{url}
\definecolor{links}{RGB}{78,102,255}
\hypersetup{
    colorlinks=true,
    linkcolor=links,
    citecolor=red,
    filecolor=links,      
    urlcolor=links,
    menucolor= links
    }
\usepackage{listings}
\usepackage{psfrag}
\usepackage{leftidx}
\usepackage{colordvi}
\usepackage{bbm}
\usepackage[caption=false]{subfig}

\newcommand{\dd}{\mathrm{d}}
\newcommand{\eE}{\mathrm{e}}
\newcommand{\ee}{\mathrm{e}}

\newcommand{\Tr}{\mathrm{Tr}}
\newcommand{\nc}{\newcommand}
\nc{\ir}{\mathrm{i}}

\begin{document}

\title{Entanglement Hamiltonian  after a local quench}

\author{Riccarda Bonsignori}
\affiliation{%
  Institute of Theoretical and Computational Physics,
  Graz University of Technology, Petersgasse 16, 8010 Graz, Austria
}%

\author{Viktor Eisler}
\affiliation{%
  Institute of Theoretical and Computational Physics,
  Graz University of Technology, Petersgasse 16, 8010 Graz, Austria
}%
\affiliation{%
  Institute of Physics, University of Graz, Universit\"atsplatz 5, 8010 Graz, Austria
}%

\begin{abstract}

We investigate the dynamics of the entanglement Hamiltonian in a system of one-dimensional free fermions, following a local joining quench of two initially disconnected half-chains in their ground states. Applying techniques of conformal field theory, we obtain a local expression where the left- and right-moving components of the energy density are associated with different weight functions. The results are then compared to numerical calculations for the hopping chain, which requires to consider a proper continuum limit of the lattice entanglement Hamiltonian, obtaining a good agreement with the field-theory prediction.

\end{abstract}


\maketitle

\section{Introduction}

Entanglement is a key concept of quantum mechanics, capturing the non-classical correlations between various parts of a system, and its characterization plays a central role at the interface of quantum information theory and many-body physics \cite{AFOV-08,CCD-09,Laflorencie-16}. For a bipartite system, all the information on entanglement between a subsystem $A$ and its complement is encoded in the reduced density matrix $\rho_A$. Entanglement measures capture the essence of this information, and have emerged as powerful tools for probing universal properties of many-body ground states \cite{VLRK-03,CC-09,ECP-10}, as well as for understanding the dynamics of quantum systems out of equilibrium \cite{CC-16,Calabrese-18}.

Despite their huge success, the various measures cannot uncover the full structure of entanglement in a given bipartition of a quantum state. However, in various situations, it is possible to get a complete description of the reduced density matrix $\rho_A  \propto \eE^{-\mathcal{H}}$ via the so-called entanglement Hamiltonian $\mathcal{H}$ (EH) \cite{DEFV-23}. The most rigorous results are available in the framework of algebraic quantum field theory (QFT), where the EH is known as the modular Hamiltonian.
Indeed, the seminal results of Bisognano and Wichmann (BW) \cite{BW-75, BW-76} identified the modular Hamiltonian of a half-infinite bipartition of the vacuum state with the generator of Lorentz boosts. The BW result holds for an arbitrary relativistic QFT, and can be generalized to other bipartitions in the presence of conformal symmetry \cite{HL-82,CHM-11, CT-16}. In the particular case of a $1+1$ dimensional conformal field theory (CFT) and a simply connected subsystem $A$, the ground-state EH can be expressed in the form
 \begin{equation}
 \label{eq:EH0}
  \mathcal{H}=\int_A \dd x \, \beta_0(x) \, T_{00}(x) ,
 \end{equation}
where $T_{00}(x)$ is the energy density of the CFT, and $\beta_0(x)$ is an appropriate weight function. In the BW case $\beta_0(x)$ is linear, while for an interval it is a parabola, with their zeros located at the boundaries of $A$. Due to the obvious thermodynamic analogy, $\beta_0(x)$ is often interpreted as a local inverse temperature \cite{WKPZV-13,ABCH-17}. However, the strict locality of the result \eqref{eq:EH0} breaks down for bipartitions with multiple intervals \cite{CH-09, LMR-10,ACHP-18} or systems with boundaries or defects \cite{MT-21a,MT-21b,RMTC-23}. 

The locality structure of the EH is also modified for lattice models with an effective low-energy CFT description.
This is most prominently observed for fermionic hopping chains, where one has direct access to the EH \cite{PE-09}, which involves long-range hopping terms \cite{EP-17,EP-18}.
Nevertheless, these can be seen as lattice artifacts and can be absorbed by considering a proper continuum limit of the EH hopping matrix \cite{ABCH-17,ETP-19}, such that the CFT result is recovered. A similar approach can be developed to match the lattice EH to the CFT results in case of multiple intervals \cite{ETP-22}, inhomogeneous fermion systems \cite{RSC-22,BE-24,BBEPV-24}, and even for bosonic theories \cite{DGT-20,JT-22,GRT-25}.

The situation becomes more involved in the out-of-equilibrium regime. The simplest protocol accessible via CFT methods is the global quench \cite{CC-05}, where the system is prepared in the ground state of a Hamiltonian, and then time evolved with another one that differs extensively. Although a global quench leads to a rapid increase in entanglement \cite{CC-05,CC-16}, the expression of the EH in the CFT framework remains local \cite{CT-16}. However, in contrast to \eqref{eq:EH0}, the result is now obtained in terms of the right- and left-moving energy densities $T(x-t)$ and $\bar T(x+t)$, respectively, with associated weight functions that differ from each other. The time evolution of the EH in a global quench setup can also be studied for hopping chains \cite{dGAT-19}, and has recently been understood in terms of the quasiparticle picture \cite{RRC-24,TRC-25}. Remarkably, the lattice result differs markedly from the CFT expectation, as the EH becomes a genuinely long-range operator. Indeed, the CFT result is only recovered assuming a linear dispersion, and for an initial state described by a thermal occupation function in terms of the modes of the post-quench Hamiltonian \cite{RRC-24}. 

In this paper we shall instead study a \emph{local quench} protocol, where initial and final Hamiltonians differ only by a local term. One of the most investigated protocols is the so-called \textit{joining quench}, where two half-chains prepared in their ground states are connected together. The dynamics of entanglement after a joining quench has been studied in the context of lattice models
 \cite{EP-07,EKPP-08,ISL-09,EP-12,CC-13,VS-17,GT-21}, quantum field theories \cite{CC-07,SD-11,Cardy-11,AB-14,WCR-15,CdG-25} and holography \cite{NNT-13,U-13,STW-19,KFNNRT-23}. Although the time evolution of the EH for a joining quench between two half-infinite chains has also been provided in the CFT framework \cite{CT-16}, its lattice implementation is still missing. The main goal of this work is to fill this gap by studying a joining quench for a finite hopping chain, as depicted in Fig.~\ref{fig:latticeLQ}. We first generalize the CFT treatment of \cite{CT-16} to the finite geometry at hand, and derive explicitly the weights associated to the right- and left-moving components of the energy density. In a next step, we obtain exact numerical results for the lattice EH and show how they can be matched with the CFT prediction via an appropriate continuum limit. In turn, for this low-energy quench protocol, the lattice and CFT results are perfectly compatible with each other, in sharp contrast to what has been found for a global quench \cite{RRC-24}.


The manuscript is organised as follows. In Sec.~\ref{sec:model}, we introduce the free-fermion hopping chain and the setup of the local joining quench. In Sec.~\ref{sec:CFT} we present the CFT derivation of the EH, 
followed by our numerical results in Sec.~\ref{sec:CL}, which are compared by an appropriate continuum limit procedure. We discuss our findings in Sec.~\ref{sec:discussion}, and provide some details of the calculations in two Appendices.

\section{The model}
\label{sec:model}

%
\begin{figure}[t]
    \centering
    \includegraphics[width=0.6\linewidth]{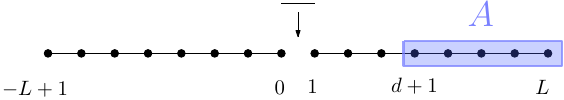}
    \caption{Geometry of the local quench. The left and right halves of the chain are joined at $t=0$.} 
    \label{fig:latticeLQ}
\end{figure}

We consider a system of free fermions hopping on a finite chain with $2L$ sites. Initially, the left and right halves of the chain are disconnected and the system is described by the Hamiltonian 
\begin{equation}
\label{eq:Hchain}
\hat{H}_0=\hat{H}_{\ell} + \hat{H}_r = -\frac{1}{2}\sum_{n=-L+1}^{-1}\left( c_n^{\dagger}c_{n+1}+c_{n+1}^{\dagger}c_n\right)
-\frac{1}{2}\sum_{n=1}^{L-1}\left( c_n^{\dagger}c_{n+1}+c_{n+1}^{\dagger}c_n\right),
\end{equation}
%
where $c^\dagger_n$ and $c_n$ are fermionic creation/annihilation operators.
We assume that the two half-chains are in their ground states at half filling. In the local quench protocol, the two half-chains are joined at time $t=0$, and the system evolves under the action of the homogeneous Hamiltonian
\begin{equation}
\label{eq:Hprimechain}
\hat{H}=-\frac{1}{2}\sum_{n=-L+1}^{L-1}\left( c_n^{\dagger}c_{n+1}+c_{n+1}^{\dagger}c_n\right).
\end{equation}
%
The local quench thus simply corresponds to replacing the hopping between sites $n=0$ and $n=1$, as illustrated in Fig.~\ref{fig:latticeLQ}. Since both the pre-quench and post-quench Hamiltonians $\hat H_0$ and $\hat H$ are quadratic in the fermions, the time evolution of the state $\ket{\psi(t)}$ is fully characterised by the two-point correlations
\begin{equation}
    C_{m,n}(t) = \bra{\psi(t)} c_m^{\dagger}c_{n} \ket{\psi(t)},
    \qquad
    \ket{\psi(t)} = \ee^{-\ir \hat H t} \ket{\psi_0}_\ell \otimes \ket{\psi_0}_r
\end{equation}
where $\ket{\psi_0}_{\ell,r}$ denotes the ground state of the left/right half-chain.

The time evolution of the correlation matrix can be obtained in the Heisenberg picture as
\begin{equation}
    C(t)=U^\dag C(0) \, U,
    \label{eq:Ct}
\end{equation}
where $U$ is the time-evolution operator expressed in the single-particle basis, with matrix elements given by
\begin{equation}
    U_{m,n}=\left[ \eE^{- \ir \hat{H} t} \right]_{m,n}=\sum_{k=1}^{2L}\Phi_k(m)\Phi_k(n)\eE^{- \ir \omega_k t}.
\end{equation}
The single-particle eigenvectors and eigenvalues of $\hat H $ are given, respectively, as
\begin{equation}
\label{eq:Phi}
    \Phi_k(m)=\sqrt{\frac{2}{2L+1}}\sin \frac{\pi k m}{2L+1}, \hspace{1cm} \omega_k=-\cos\frac{\pi k}{2L+1},
\end{equation}
where $k=1, \dots,2L$. Furthermore, the initial correlation matrix has the block form
\begin{equation}
\label{eq:C0}
C(0)=\begin{pmatrix}
C_{0}& 0 \\
0 & C_{0}
\end{pmatrix},
\end{equation}
where each block corresponds to the ground-state correlation matrix of a half-filled open chain of length $L$, with matrix elements \cite{FC-11} 
\begin{equation}
 [C_{0}]_{mn}=\frac{1}{2(L+1)}\left[ \frac{\sin (\frac{\pi}{2}(n-m))}{\sin \left(\frac{\pi}{2(L+1)}(n-m) \right)} -\frac{\sin (\frac{\pi}{2}(n+m))}{\sin \left(\frac{\pi}{2(L+1)}(n+m) \right)}\right]. 
 \label{eq:Cgs}
\end{equation}

We are interested in the EH associated to a subsystem $ A = [d+1, L] $, located at distance $d$ from the quench point within the right half-chain and extending to its right boundary. In fact, the reduced density matrix $\rho_A (t) = \Tr_B \ket{\psi(t)}\bra{\psi(t)}$ at time $t$ can be written in the exponential form \cite{Peschel03}
%
\begin{equation}
\label{eq:EH}
\rho_A(t)=\frac{\eE^{-\mathcal{H}(t)}}{\mathcal{Z}}, \quad \quad \mathcal{H}(t)=\sum_{k}H_{i,j}(t) c_i^{\dagger}c_j,
\end{equation}
where the EH is a free-fermion operator $\mathcal{H}(t)$. Moreover, the associated hopping matrix $H(t)$ can be related to the reduced correlation matrix $C_A(t)$, defined as the restriction of \eqref{eq:Ct} to the indices $i,j \in A$, as
%
\begin{equation}
H(t)=\mathrm{ln}\left[\frac{1-C_A(t)}{C_A(t)}\right].
\label{eq:Ht}
\end{equation}
Hence, evaluating the correlation matrix $C_A(t)$ numerically via the expressions \eqref{eq:Ct}-\eqref{eq:Cgs} gives access to the EH matrix.
However, one faces the usual problem encountered also in ground-state EH calculations, namely the accumulation of the eigenvalues of $C_A(t)$ exponentially close to zero and one. Indeed, these eigenvalues deliver the largest contribution in \eqref{eq:Ht} and thus have to be resolved, which requires very high precision numerics. Before embarking on the numerical analysis, we first present a CFT treatment of the problem.

\section{CFT formulation}
\label{sec:CFT}

In this section, we derive the EH for the joining local quench protocol in the framework of CFT. We will follow closely the treatment presented in \cite{CT-16} for a local quench in an infinite system ($L \to \infty$), and generalize it to finite $L$ with open boundary conditions.
The first step is to construct the path-integral representation of the time-evolved density matrix $\rho(t)$ \cite{SD-11}. Its matrix elements between field configurations $\ket{\varphi}$ and $\ket{\varphi'}$ are expressed as

%
\begin{equation}
\label{eq:rho}
\langle \varphi''|\rho(t)|\varphi'\rangle = Z^{-1} \langle \varphi''|\eE^{-\ir t \hat H} \eE^{-\lambda \hat H}|\varphi_{\ell} \otimes \varphi_r \rangle\langle \varphi_{\ell}  \otimes \varphi_r  | \eE^{\ir t \hat H} \eE^{-\lambda \hat H} |\varphi'\rangle
\end{equation}
where $\ket{\varphi_{\ell}  \otimes \varphi_r}$ denotes the initial tensor product of the two ground states on the half-chains and $Z$ is a normalization. The damping factors $\eE^{-\lambda \hat H}$ are introduced to make the path integral absolutely convergent, and we will be interested in the regime $t \gg \lambda$. The expression \eqref{eq:rho} can be turned into an Euclidean path integral by introducing the imaginary time $\tau =it$. The time evolution then corresponds to path integration of width $\lambda \pm \tau$ along the imaginary time direction, from the corresponding boundary conditions $\ket{\varphi_{\ell}  \otimes \varphi_r}$ and $\bra{\varphi_{\ell}  \otimes \varphi_r}$, respectively. Moreover, these boundary states can themselves be expressed as path integrals using
%
\begin{equation}
    |\varphi_{\ell}\otimes\varphi_{r}\rangle =\lim_{\beta \rightarrow \infty}\frac{\eE^{-\beta(\hat H_{\ell}+\hat H_{r})}\eE^{\beta E_0}|s\rangle}{\langle \varphi_{\ell}\otimes\varphi_{r}|s\rangle}.
    \label{eq:gscft}
\end{equation}
Indeed, in the limit $\beta \to \infty$, the arbitrary initial state $\ket{s}$ is projected towards the ground state, assuming they have a nonzero overlap $\langle \varphi_{\ell}\otimes\varphi_{r}| s \rangle \neq 0$, and $E_0$ is the ground state energy of $\hat H_{\ell}+\hat H_r$.

\begin{figure}[t]
    \centering
    \includegraphics[width=0.9\linewidth]{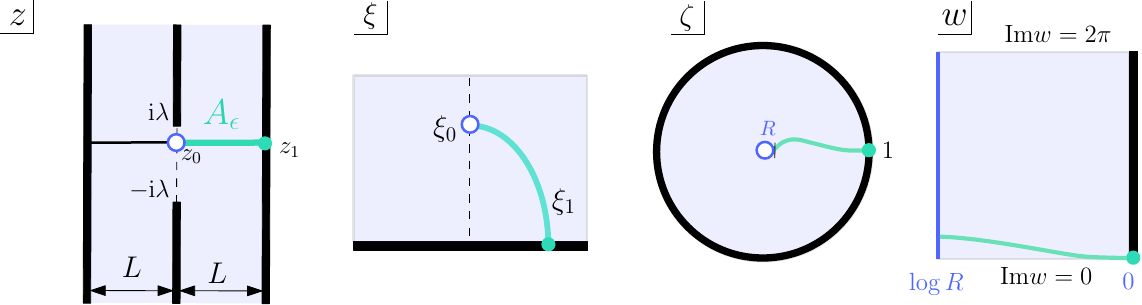}
    \caption{Mapping from the double-pants geometry to the logarithmic image of the annulus.} 
    \label{fig:ConformalMapQuench}
\end{figure}

Putting the pieces together, the local quench has the path-integral representation shown in the first panel of Fig.~\ref{fig:ConformalMapQuench}, which was dubbed the double-pants geometry \cite{SD-11}. The legs of the lower pant with $- \infty < \mathrm{Im}(z) < - \lambda$ correspond to the imaginary time evolution creating the initial state in \eqref{eq:gscft}, while those of the upper pant with $\lambda < \mathrm{Im}(z) < \infty$ yield $\bra{\varphi_{\ell}  \otimes \varphi_r}$. The thick lines represent the physical boundaries of the system, with identical boundary conditions. The time evolution after the quench then takes place in the center with $- \lambda < \mathrm{Im}(z) < \lambda$, and the matrix element in \eqref{eq:rho} is thus given by a path integral with a slit along $\mathrm{Im}(z)=\tau$, which is depicted by the thin horizontal black line in Fig.~\ref{fig:ConformalMapQuench}. For the reduced density matrix the field configurations in part $B=[-L,x_0]$ are integrated over, and thus the cut extends along $C=\{z=x+\ir \tau, x_0<x<L\}$, which is shown by the thick cyan line.
The regularization of the path integral is obtained by cutting out a small disk of radius $\epsilon$ around the entangling point $z_0=x_0+\ir \tau$, while the other endpoint is denoted $z_1=L+\ir\tau$. Note that $\tau$ has to be  considered real during the calculation, and only at the end analytically continued to $\tau \to \ir t$.

%


In order to obtain the EH, one has to map the regularized double pants geometry into an annulus \cite{CT-16}. In the first step, the double pants is mapped to the upper half-plane (UHP) via the conformal mapping \cite{SD-11,KFNNRT-23}
\begin{equation}
\label{eq:xi}
    \xi(z)=\ir \sqrt{\frac{\sin \left(\frac{\pi}{2L}(\ir \lambda+z) \right)}{\sin \left(\frac{\pi}{2L}(\ir \lambda-z) \right)}},
\end{equation}
which sends the boundaries of the strip as well as the cut between the legs to the real axis. The regularised slit $C_{\epsilon}=[z_0+\epsilon, z_1]$ is transformed into a curve on the UHP, with extrema  $\xi_0=\xi(z_0)$ and $\xi_1=\xi(z_1)$, as shown in the second panel of Fig. \ref{fig:ConformalMapQuench}.  We subsequently map the UHP to an annulus via the mapping
\begin{equation}
\label{eq:zeta}
    \zeta(z)= \left( \frac{\xi_1-\bar{\xi}_{0}}{\xi_1-\xi_0}\right)\frac{\xi(z)-\xi_0}{\xi(z)-\bar{\xi}_{0}},
\end{equation}
such that $\zeta(z_0)=0$, i.e. the entangling point is mapped to the center of the disk with an inner radius $R \propto \epsilon$, and $\zeta(z_1)=1$, as shown in the third panel of Fig. \ref{fig:ConformalMapQuench}. However, unlike in equilibrium situations, the image of the regularized interval does not correspond to a radius in the annulus. Finally, taking the logarithm of this transformation,
\begin{equation}
\label{eq:wmap}
    w(z)=\log\left[ \left( \frac{\xi_1-\bar{\xi}_{0}}{\xi_1-\xi_0}\right)\frac{\xi(z)-\xi_0}{\xi(z)-\bar{\xi}_{0}}\right],
\end{equation}
the punctured disk is mapped to the strip in the $w$-plane, as shown in the fourth panel of Fig.\ref{fig:ConformalMapQuench}, where $\mathrm{Im}(w)=0$ is identified with $\mathrm{Im}(w)=2\pi$.

In the final strip geometry, the EH is given by the integral of the generator of translations in the imaginary time direction. However, since the image $w(C_\epsilon)$ is not a constant time-slice, the result must be written using the holomorphic/antiholomorphic components of the energy-momentum tensor \cite{CT-16}
%
\begin{equation}
\mathcal{H}=2 \pi\int_{w(C_{\epsilon})}T(w)\dd w+2 \pi\int_{\overline{w(C_{\epsilon})}}\bar{T}(\bar{w})\dd \bar{w}.
\end{equation}
The result can now be transformed back into the original double-pants geometry as
\begin{equation}
\mathcal{H}=\int_{C_{\epsilon}} \beta(z)T(z)\dd z+  \int_{\bar{C}_{\epsilon}} \bar{\beta}(\bar z)\bar{T}(\bar{z})\dd \bar{z},
\end{equation}
where the coefficients 
\begin{equation}
    \beta(z) = \frac{2\pi}{w'(z)}, \qquad
    \bar \beta(\bar z)= \frac{2\pi}{\overline{w'(z)}}
    \label{eq:betaz}
\end{equation}
are obtained by combining the transformation properties of the energy-momentum tensor components and the Jacobian factor \cite{CT-16}. In particular, using \eqref{eq:wmap} we get
%
%
%
\begin{equation}
\label{eq:wprime}
 \beta(z) =2 \pi \frac{(\xi(z)-\xi_0)(\xi(z)-\Bar{\xi}_{0})}{(\xi_0-\bar{\xi}_{0})\xi'(z)}.
\end{equation}

Finally, one has to substitute for the complex coordinate $z=x+i\tau$ and take the analytic continuation $\tau \to \ir t$. The variables thus become $z=x-t$ and $\bar z = x+t$ and the integrals along the contours $C_\epsilon$ and $\bar C_\epsilon$ can be written as a spatial integral along the subsystem $A_\epsilon$ as
\begin{equation}
\label{eq:EHCFT}
        \mathcal{H} = \int_{A_\epsilon} \beta(x,t) T(x-t)\dd x +\int_{A_\epsilon} \bar{\beta}(x,t)\bar{T}(x+t)\dd x .
\end{equation}
In turn, one can interpret the result for the EH as an effective thermal expression, where the right-moving $T(x-t)$ and left-moving $\bar T(x+t)$ components of the energy density are associated with different local inverse temperatures $\beta(x,t)$ and $\bar\beta(x,t)$. Note that the respective inverse temperatures are not only functions of $x \mp t$, as the time enters the definition \eqref{eq:wprime} via $\xi_0$ in a nontrivial way.

\subsection{EH for the half-chain}

We can now proceed to derive explicit expressions for $\beta(x,t)$ and $\bar\beta(x,t)$. We start by considering the half-chain partition, $x_0=0$, where the entangling point and the quench location coincide. In this case, the image of $z_0$ in the UHP is purely imaginary,
\begin{equation}
\xi_0 =\xi(\ir \tau)=\ir \sqrt{\frac{\sinh\frac{\pi}{2L}(\lambda+\tau)}{\sinh\frac{\pi}{2L}(\lambda-\tau)}},
\end{equation}
and the numerator of \eqref{eq:wprime} can be simplified as
\begin{equation}
(\xi(z)-\xi_0)(\xi(z)-\Bar{\xi}_{0})=
    -\ir \frac{\sin\frac{\pi}{2L}(z-\ir\tau) \sinh\frac{\pi \lambda}{L}}{\sin\frac{\pi}{2L}(\ir\lambda-z)\sin\frac{\pi}{2L}(\ir\lambda-\ir\tau)}.
\end{equation}
For the denominator we also need the derivative
\begin{equation}
\label{eq:xiprime}
    \xi'(z)=-\frac{\pi}{4L}\frac{\sinh \left(\frac{\pi}{L}\lambda \right)}{\sin^2\left( \frac{\pi}{2L}(\ir \lambda-z)\right)}\sqrt{\frac{\sin \left( \frac{\pi}{2L}(\ir \lambda - z)\right)}{\sin \left( \frac{\pi}{2L}(\ir  \lambda + z)\right)}}.
\end{equation}
Plugging in all these expressions into \eqref{eq:wprime} and using trigonometric identities, we arrive at
\begin{equation}
 \beta(z)=4L\sin \left[ \frac{\pi}{2L}(z-\ir \tau)\right]\sqrt{\frac{\cosh \frac{\pi \lambda}{L}-\cos \frac{\pi z}{L}}{\cosh \frac{\pi \lambda}{L}-\cosh \frac{\pi \tau}{L}}}.
\end{equation}
Finally, we need to substitute $z=x+\ir \tau$ and take the analytic continuation $\tau\rightarrow \ir t$, which yields
\begin{equation}
       \beta(x,t)=  4L \sin  \frac{\pi x}{2L} \sqrt{\frac{\cosh \frac{\pi \lambda}{L}- \cos \frac{\pi }{L}(x-t)}{\cosh \frac{\pi \lambda}{L}- \cos \frac{\pi t}{L}}}.
       \label{eq:betaxt}
\end{equation}
It is easy to see that the result for $\bar\beta(x,t)$ follows by exchanging $x-t$ with $x+t$. The expressions can be further simplified in the regime $t \gg \lambda$. Indeed, one can then set $\lambda=0$ in \eqref{eq:betaxt} to obtain
\begin{equation}
\label{eq:betabetabarx0}
\beta(x,t)=4L\left|\frac{\sin\frac{\pi x}{2L}\sin \frac{\pi}{2L}(x-t)}{\sin \frac{\pi t}{2L}}\right| , \qquad \bar{\beta}(x,t)=4L\left|\frac{\sin\frac{\pi x}{2L}\sin \frac{\pi}{2L}(x+t)}{\sin \frac{\pi t}{2L}}\right| .
\end{equation}

\begin{figure}[t]
    \hspace*{-0.2cm}
    \subfloat{\includegraphics[width=0.45\textwidth]{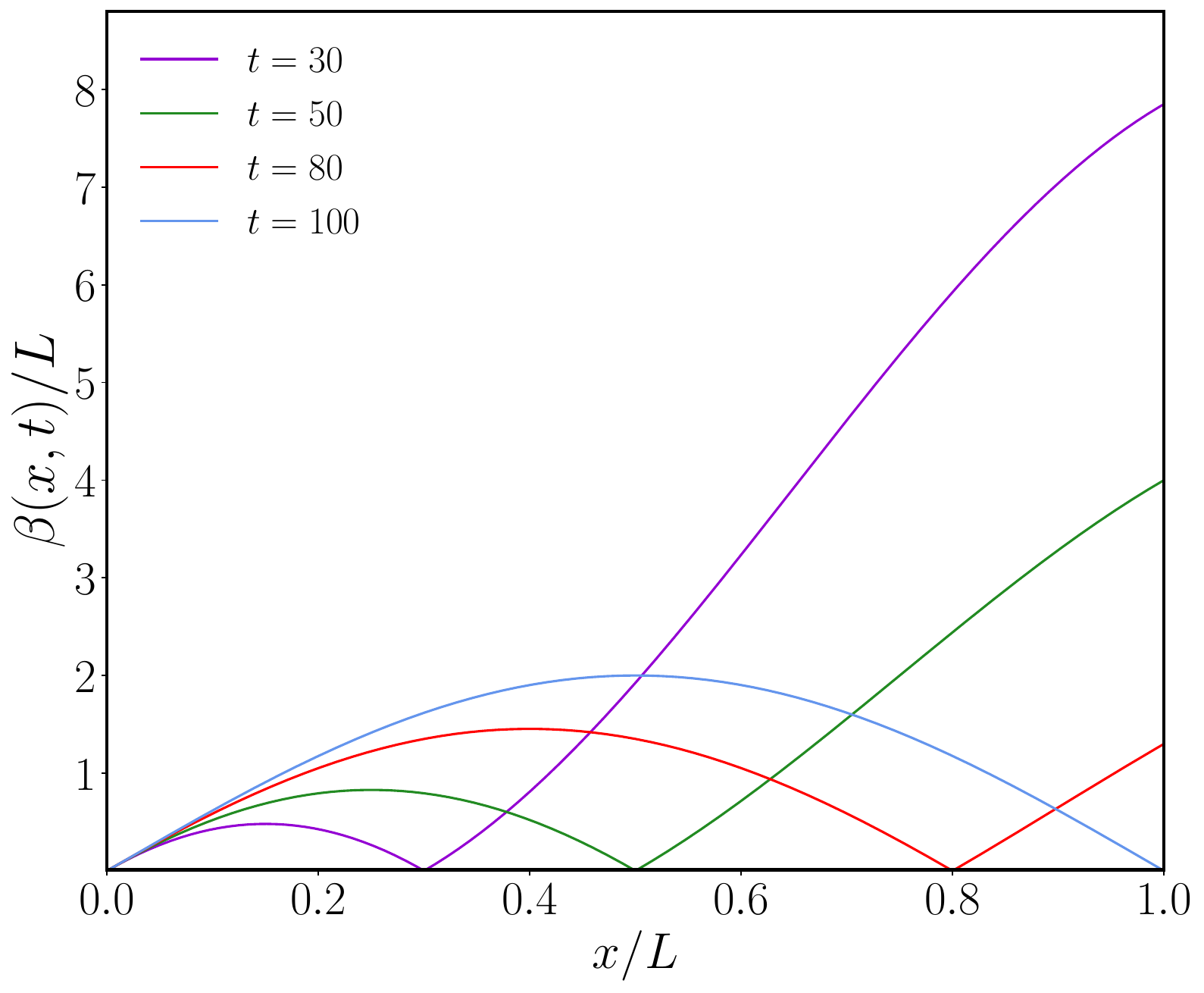}}
    \hspace{0.5cm}\subfloat{\includegraphics[width=0.45\textwidth]{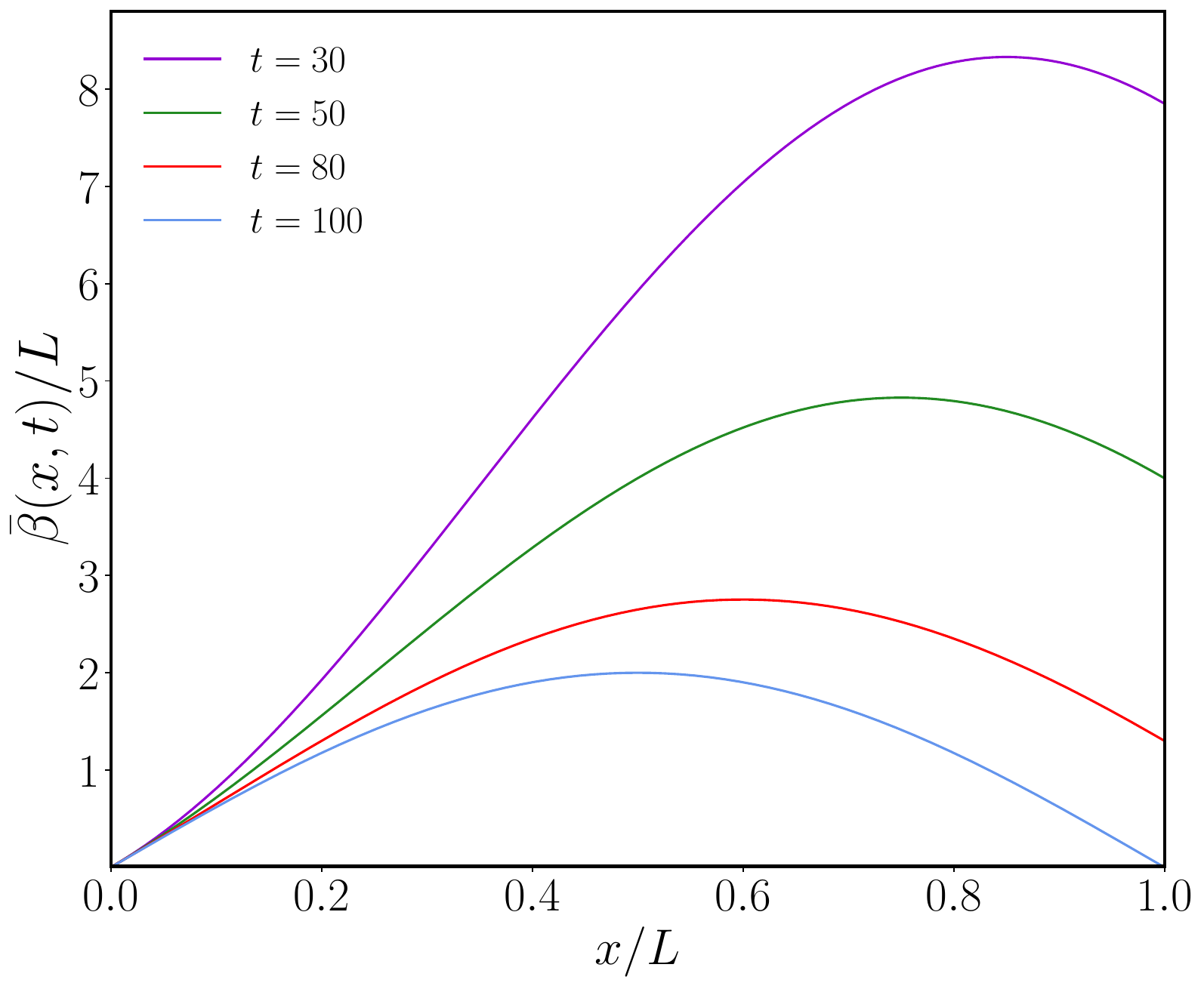}}
   \caption{Inverse temperatures $\beta(x,t)/L$ (left) and $\bar{\beta}(x,t)/L$ (right) for $L=100$ and different times, in the limit $\lambda\to 0$ as given by \eqref{eq:betabetabarx0}.}
     \label{fig:betabetabarx00}
\end{figure}

In Fig.~\ref{fig:betabetabarx00}, we show the plots of $\beta(x,t)/L$ (left) and $\bar{\beta}(x,t)/L$ (right) as functions of the rescaled position $x/L$, for $L=100$ and various times $t \le L$. 
Clearly, the inverse temperature for the right-moving part displays two zeros, one at the subsystem boundary $x=0$ and another one at $x = t$, which is the location of the front that propagates ballistically due to the perturbation induced by the quench. In fact, the presence of this propagating front was observed numerically in the eigenvectors of the correlation matrix in the $L\to \infty$ limit of the quench \cite{EP-07}. The result \eqref{eq:betabetabarx0} now makes it clear, that the right-moving front creates a new entangling point within the subsystem, splitting it in two parts.
In contrast, the left-moving front creates an effective boundary at $x=-t$, which lies outside $A$ and the corresponding inverse temperature $\bar{\beta}(x,t)$ has a single zero for $t<L$. For $t=L$, however, as the right-moving front is reflected from the boundary, $\bar{\beta}(x,t)$ develops another zero at $x=L$, and the inverse temperatures become equal, $\beta(x,L)=\bar\beta(x,L)$.
For times $L<t<2L$, the front is located at $x=2L-t$ after the reflection, and one finds from \eqref{eq:betabetabarx0} the symmetry property $\bar{\beta}(x,t)=\beta(x,2L-t)$ and $\beta(x,t)=\bar\beta(x,2L-t)$. In other words, the roles of the right- and left-moving inverse temperatures are simply interchanged after the front is reflected from the chain edge.


It is instructive to check the $L\to \infty$ limit of our result
\begin{equation}
        \beta(x,t)=2 \pi\frac{x|x-t|}{t} , \qquad 
        \bar{\beta}(x,t)=2\pi \frac{x(x+t)}{t},
\end{equation}
which is precisely the result obtained in \cite{CC-16}. The inverse temperatures have then a parabolic form, and 
$\beta(x,t)$ is identical to the ground-state result of an interval $[0,t]$ in an infinite chain.


\subsection{EH for arbitrary $x_0$}

We now consider an arbitrary partition where the boundary point $x_0$ between the two subsystems does not coincide with the quench location $x=0$. Without loss of generality, we will assume $x_0>0$. The computation of $\beta(z)$ for this decentered segment is slightly more complicated, due to the proper handling of the branch cuts in the mapping \eqref{eq:xi}. Indeed, it is necessary to keep track of the additional sign factor which could appear depending on the argument of the square root. This is discussed in detail in Appendix \ref{app:A}, and we only report the result here.
For times $t<L$ before the first reflection we find 
%
\begin{equation}
\label{eq:betadec}
\beta(x,t)= \left\{ 
    \begin{aligned}
   &2L\frac{\cos\frac{\pi x_0}{L}-\cos\frac{\pi x}{L}}{\sin\frac{\pi x_0}{L}} \\
   & 4L \frac{\sin \frac{\pi}{2L}(x-x_0)\sin \frac{\pi}{2L}(t-x)}{\sin \frac{\pi}{2L}(t-x_0)} \\
   & 4L \frac{\sin \frac{\pi}{2L}(x+x_0)\sin \frac{\pi}{2L}(x-t)}{\sin \frac{\pi}{2L}(t+x_0)}
    \end{aligned}
    \right.
    \quad
    \bar{\beta}(x,t)=\left\{ 
    \begin{aligned}
  & 2L\frac{\cos\frac{\pi x_0}{L}-\cos\frac{\pi x}{L}}{\sin\frac{\pi x_0}{L}} & \quad t<x_0<x \\
   & 4L\frac{\sin \frac{\pi}{2L}(x-x_0)\sin \frac{\pi}{2L}(x+t)}{\sin \frac{\pi}{2L}(t+x_0)}
    & \quad x_0<x<t \\
   & 4L \frac{\sin \frac{\pi}{2L}(x-x_0)\sin \frac{\pi}{2L}(x+t)}{\sin \frac{\pi}{2L}(t+x_0)}
    & \quad x_0<t<x
    \end{aligned}
    \right. 
\end{equation}

The result can be discussed as follows. For short times $t<x_0$, when the front propagating from the initial cut has not yet reached the subsystem, we recover the ground-state result of a segment $[x_0,L]$ in a finite chain of length $L$
for both inverse temperatures. When the front reaches the subsystem, the inverse temperatures become time dependent and one has to distinguish between two regimes. In particular, the expression of $\beta(x,t)$ changes on the two sides of the front, $x<t$ and $x>t$, respectively. Interestingly, in the latter case the location $-x_0$ of the reflected subsystem boundary enters. Note that the expression of $\bar\beta(x,t)$ remains unchanged. Qualitatively, the inverse temperatures are very similar to those plotted in Fig.~\ref{fig:betabetabarx00},
and are not shown here. Finally, we remark that for times $L<t<2L-x_0$ one has again the symmetry property $\bar{\beta}(x,t)=|\beta(x,2L-t)|$ and $\beta(x,t)=|\bar\beta(x,2L-t)|$.

%


%

\subsection{Entanglement entropy}

As already pointed out in \cite{CC-16} for the case $L \to \infty$, the different number of roots of $\beta(x,t)$ and $\bar\beta(x,t)$ for $x>0$ immediately implies that the contribution to the entanglement entropy from the right-movers is three times the one from the left-movers. In the following we shall compute the respective contributions for finite $L$,
focusing on $x_0=0$. We make use of the linear dependence of the entropy density $s=\frac{\pi c T}{3}$ on the temperature in 1+1D CFT. In the quench problem, the temperature is a slowly varying function in space and differs for the left/right-moving component of the energy density. This suggests to express the entropy as a sum of contributions coming from the right- and the left-movers in the form
\begin{equation}
\label{eq:EE}
    S \equiv S_R+S_L= \frac{\pi c}{6} \int_{A_\epsilon} \frac{\dd x}{\beta (x,t)}+\frac{\pi c}{6}\int_{A_\epsilon} \frac{\dd x}{\bar{\beta} (x,t)}
\end{equation}
In other words, the chiral components contribute with a halved prefactor and the effective local temperature is integrated over within the regularized subsystem. In fact, working directly in the limit $\lambda \to 0$, the right-moving temperature diverges also at $x=t$, and an additional regularization is necessary.  Introducing $A_{\epsilon,\lambda}=[\epsilon,t-\lambda]\cup[t+\lambda,L]$ and using \eqref{eq:betabetabarx0}, the  contribution can be evaluated as
\begin{equation}
    S_R = \frac{\pi c}{6}\int_{A_{\epsilon,\lambda}} \frac{\dd x}{\beta(x,t)}
    =\frac{c}{12}\log \Bigg|\frac{\sin\frac{\pi(t-\lambda)}{2L} \sin\frac{\pi(t-\epsilon)}{2L}}{\sin\frac{\pi\lambda}{2L} \sin \frac{\pi\epsilon}{2L}}\Bigg|+\frac{c}{12}\log \Bigg|\frac{\sin \frac{\pi (t+\lambda)}{2L}\cos \frac{\pi t}{2L}}{\sin \frac{\pi \lambda}{2L}} \Bigg|,
\end{equation}
For the left-moving component there is no need to split the interval and one obtains
\begin{equation}
        S_L=
        \frac{\pi c}{6}\int_{A_\epsilon} \frac{\dd x}{\bar\beta(x,t)}=\frac{c}{12}\log \Bigg| \frac{\sin \frac{\pi(t+\epsilon)}{2L}}{\sin \frac{\pi\epsilon}{2L}\cos \frac{\pi t}{2L}}\Bigg|.
\end{equation}

In turn, in the limit $\lambda,\epsilon \ll 1$ we observe that $S_R$ has a tripled contribution with the factor $\sin \frac{\pi t}{2L}$ inside the logarithm, as compared to $S_L$. However, there is an additional factor $\cos \frac{\pi t}{2L}$ that appears with different signs in the two entropies. Summing the two contributions and expanding the factors with the regulators, one arrives at
\begin{equation}
    S
    \simeq\frac{c}{3}\log \Bigg|\frac{2L}{\pi \sqrt{\lambda \epsilon}}\sin \frac{\pi t}{2L}\Bigg|,
\end{equation}
which is precisely the result found in \cite{SD-11}.

%
%
%

%
%
%

\section{Lattice results}
\label{sec:CL}

With the CFT result at hand, we now proceed to the numerical calculation of the EH for the hopping chain introduced in Section \ref{sec:model}. Although the CFT treatment yields a perfectly local expression, this is generically not expected for lattice problems. Indeed, even in the simplest ground-state setting it was shown that the EH contains long-range hopping terms \cite{EP-17}. These are artifacts of the presence of the lattice, and for a proper comparison with CFT, one needs to perform a continuum limit \cite{ABCH-17,ETP-19}. 

Here we derive this continuum limit for the local quench, where the EH can be rewritten in the form
\begin{equation}
\label{eq:EHCL}
    \mathcal{H}=\sum_j t_0(j)c_j^{\dagger}c_j+\sum_j\sum_{r\geq 1}\left[ t_r(j+r/2)  \left( c_j^{\dagger}c_{j+r}+c_{j+r}^{\dagger}c_j\right)+ \ir s_r(j+r/2)\left( c_j^{\dagger }c_{j+r}-c_{j+r}^{\dagger}c_j\right)\right].
\end{equation}
Comparing to \eqref{eq:EH}, we can identify the hopping amplitudes as 
\begin{equation}
    t_r(j+r/2) = \mathrm{Re\,} {H_{j,j+r}(t)}, \qquad
    s_r(j+r/2) = \mathrm{Im\,} {H_{j,j+r}(t)},
\end{equation}
such that they contain the EH matrix elements along the diagonals. These are assumed to be slowly varying functions of $j$, and it is useful to define their arguments as the midpoint of the corresponding hopping. Obviously, the only difference with respect to the ground-state setting is the presence of imaginary matrix elements, which are associated to the presence of currents in the quench problem.

The continuum limit is introduced via the usual substitution
\begin{equation}
    c_j \quad \rightarrow \quad \sqrt{a}\left( \eE^{\ir q_Fx}\psi(z)+\eE^{-\ir q_F x}\bar{\psi}(\bar z)\right),
\end{equation}
where $x=j\,a$ is a continuous variable in terms of the lattice spacing $a$, and the right/left moving fields $\psi$ and $\bar{\psi}$ depend on the spacetime coordinates only via $z=x-t$ and $\bar{z}=x+t$, respectively. The phase factors $\eE^{\pm\ir q_F x}$ account for the rapid lattice oscillations due to the finite Fermi points at $\pm q_F a$, such that the fields are well defined in the continuum limit. As in the equilibrium case, the derivation consists of expanding the operators $c_{j+r}$ to lowest order in the lattice spacing, and is summarized in Appendix \ref{app:CL}. The resulting expression has the form
\begin{equation}
\label{eq:EHcontinuumlimit}
    \mathcal{H}=\int_A \dd x \left[ \beta_0(x,t) T_{00}(z,\bar z)-\beta_1(x,t)T_{01}(z,\bar z) + \mu_0(x,t)\rho(z,\bar z)-\mu_1(x,t)j(z,\bar z)\right],
\end{equation}
where the components of the stress-energy tensor for the Dirac fermion are given by 
\begin{equation}
\label{eq:TTensor}
    \begin{split}
        T_{00}(z,\bar z)&=T(z) + \bar T(\bar z) = \frac{1}{2}[\psi^{\dagger}(z)(-\ir \partial_z)\psi(z)-\bar{\psi}^{\dagger}(\bar{z})(-\ir \partial_{\bar z})\bar{\psi}(\bar{z})+\mathrm{h.c.}],\\
         T_{01}(z,\bar z)&=T(z) - \bar T(\bar z) = \frac{1}{2}[\psi^{\dagger}(z)(-\ir \partial_z)\psi(z)+\bar{\psi}^{\dagger}(\bar{z})(-\ir \partial_{\bar z})\bar{\psi}(\bar{z})+\mathrm{h.c.}].
    \end{split}
\end{equation}
whereas the density and current of the $U(1)$ charge read
\begin{equation}
    \label{eq:current}\rho(z,\bar z)=\psi^{\dagger}(z)\psi(z)+\bar{\psi}^{\dagger}(\bar z)\bar{\psi}(\bar z), \quad \quad j(z,\bar z)=\psi^{\dagger}(z)\psi(z)-\bar{\psi}^{\dagger}(\bar z)\bar{\psi}(\bar z).
\end{equation}
The weights corresponding to each of these operators follow as
\begin{equation}
\label{eq:multipliers}
    \begin{split}
        \beta_0(x,t)&=-2 a \sum_{r=1}^{\infty}r \sin(r q_F a)t_r(x) \qquad \mu_0(x,t)=t_0+2 \sum_{r=1}^{\infty}\cos(r q_F a)t_r(x), \\ \beta_1(x,t)&=2a \sum_{r=1}^{\infty}r \cos(r q_Fa)s_r(x)
         \qquad \mu_1(x,t)=2a\sum_{r=1}^{\infty} \sin(r q_F a)s_r(x).
    \end{split}
\end{equation}

The continuum limit thus yields a local expression for $\mathcal{H}$, and the weight factors are obtained by summing over the diagonals of the EH matrix calculated on the lattice. The only information that enters about the \emph{physical} Hamiltonian is the Fermi momentum. In particular, for the local quench at half filling one has $q_F a=\pi/2$, and the structure further simplifies. Indeed, due to the checkerboard structure of the initial correlation matrix in \eqref{eq:Cgs}, the time-evolved matrix $C(t)$ as well as $H(t)$ have a special pattern, with 
the diagonals being either purely real or imaginary, 
$t_{2p}(x)\equiv 0$ and $s_{2p+1}(x)\equiv 0$. This implies $\mu_0(x,t)\equiv 0$ and $\mu_1(x,t)\equiv 0$, and the nonvanishing weights read
\begin{equation}
\label{eq:CLhf}
    \beta_0(x,t)= -2a \sum_{p=0}^{\infty}(2p+1)(-1)^pt_{2p+1}(x), \quad \quad \beta_1(x,t)= 2a \sum_{p=1}^{\infty} (2p) (-1)^ps_{2p}(x).
\end{equation}
In this case, the numerical evaluation of the sums is straightforward, as it simply requires summing the matrix elements of $H(t)$ along the anti-diagonals. The number of nonvanishing entries in the sum thus depends on $x=ja$ and is limited by the size of the subsystem. The continuum limit amounts to sending $a \to 0$ and $L \to \infty$, by fixing their product as $La=1$. Setting also $da=x_0$, one has $x_0<x<1$, such that in the continuum setup the half-chain has unit length. Finally, writing the stress-tensor components in terms of the holomorphic/anti-holomorphic parts as in \eqref{eq:TTensor}, the expression \eqref{eq:EHcontinuumlimit} can be matched to the CFT prediction \eqref{eq:EHCFT} as
\begin{equation}
\label{eq:beta01}
    \beta_0(x,t)=\frac{\beta(x,t)+\bar{\beta}(x,t)}{2}, \qquad
    \beta_1(x,t)=\frac{\bar\beta(x,t)-\beta(x,t)}{2}.
\end{equation}
%

\begin{figure}[t]
    \subfloat{\includegraphics[width=0.43\textwidth]{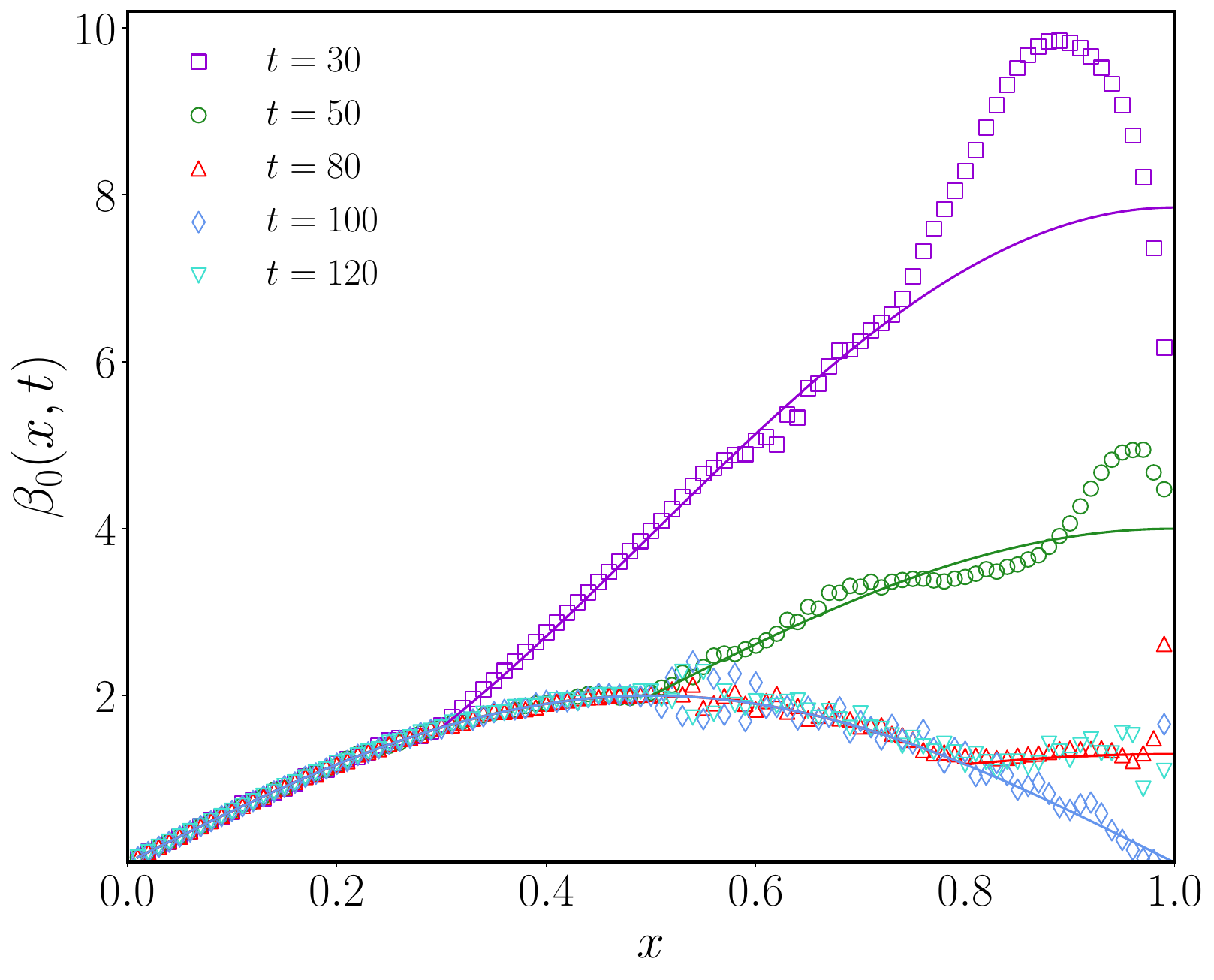}}
    \hspace{0.5cm}
\subfloat{\includegraphics[width=0.45\textwidth]{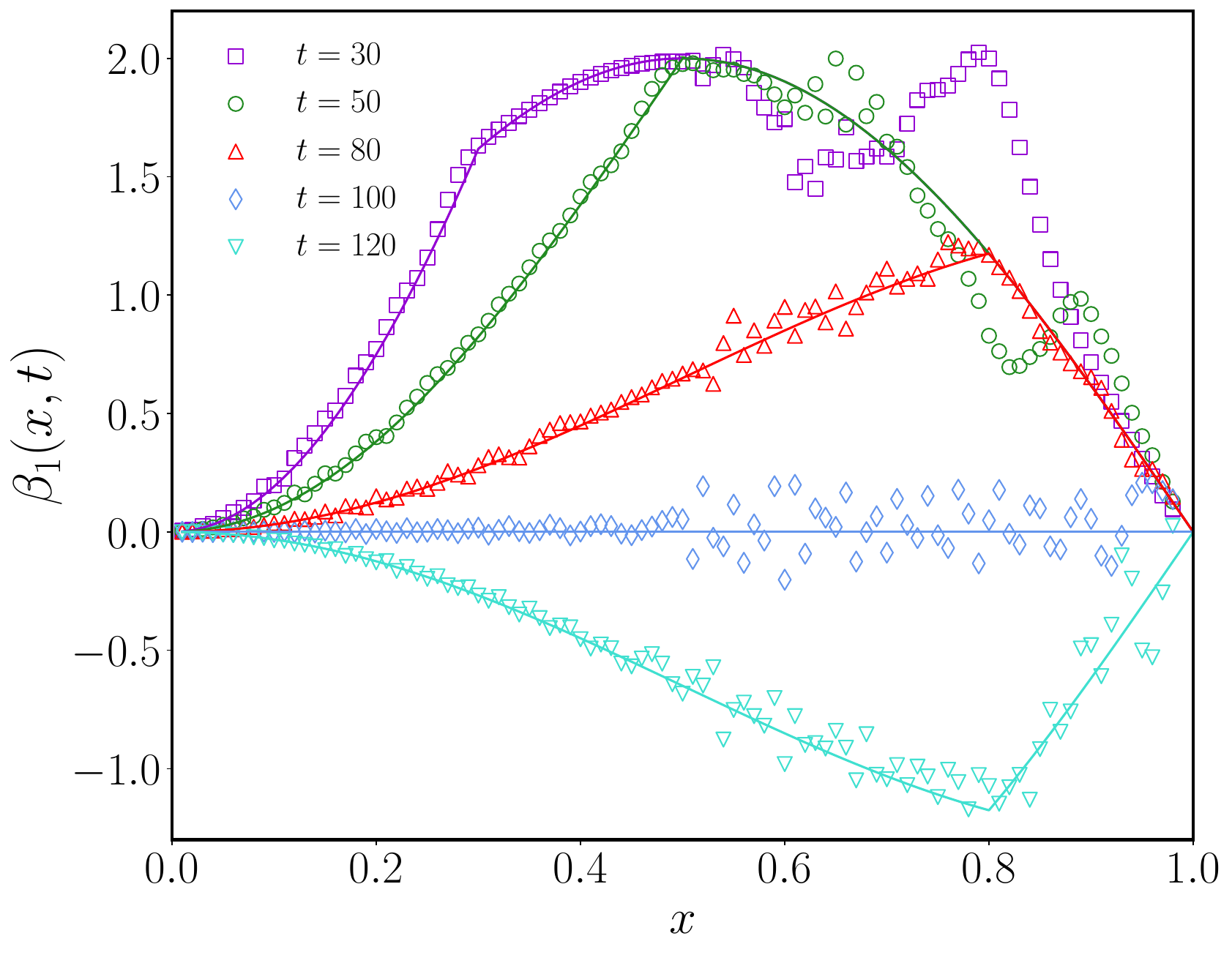}}
   \caption{Spatial profiles of the weights $\beta_0(x,t)$ (left) and $\beta_1(x,t)$ (right) as a function of $x=j/L$ and at different times $t$, for a half-chain of size $L=100$. The symbols represent the numerical data obtained from \eqref{eq:CLhf} with $a=1/L$, and the solid lines are the corresponding CFT predictions \eqref{eq:beta01} with \eqref{eq:betabetabarx0}. }
     \label{fig:betaCLx0}
\end{figure}

Our numerical results for the continuum limit of the half-chain EH of size $L=100$ are shown in Fig.~\ref{fig:betaCLx0}. The agreement between the data and the CFT prediction \eqref{eq:beta01} is remarkable, except for some oscillations in the regime $j>t$. These deviations are rather strong for small $t$, but diminish for larger times. Interestingly, the weight of the energy density becomes independent of time in the regime $x<t/L$ already visited by the front, and is given by
\begin{equation}
    \beta_0(x,t) = 2\sin (\pi x).
\end{equation}
This can be seen from the overlap of the curves on the left of Fig.~\ref{fig:betaCLx0}. A similar behaviour is observed for the weight $\beta_1(x,t)$ of the momentum density in the regime $x>t$, shown on the right of the figure. In contrast, $\beta_1(x,t)$ has a quadratic increase for $x\to 0$, with the $x<t$ part of the profile flattening out as time increases.
In particular, for $t=L$ the CFT prediction is $\beta_1(x,t)=0$, which is indeed supported by the numerical data, up to some oscillations. Furthermore, after the reflection from the chain boundary, the symmetry properties imply $\beta_0(x,2L-t)=\beta_0(x,t)$ and $\beta_1(x,2L-t)=-\beta_1(x,t)$, which is indeed visible in the data for $t=80$ and $t=120$.

\begin{figure}[t]
    \subfloat{\includegraphics[width=0.45\textwidth]{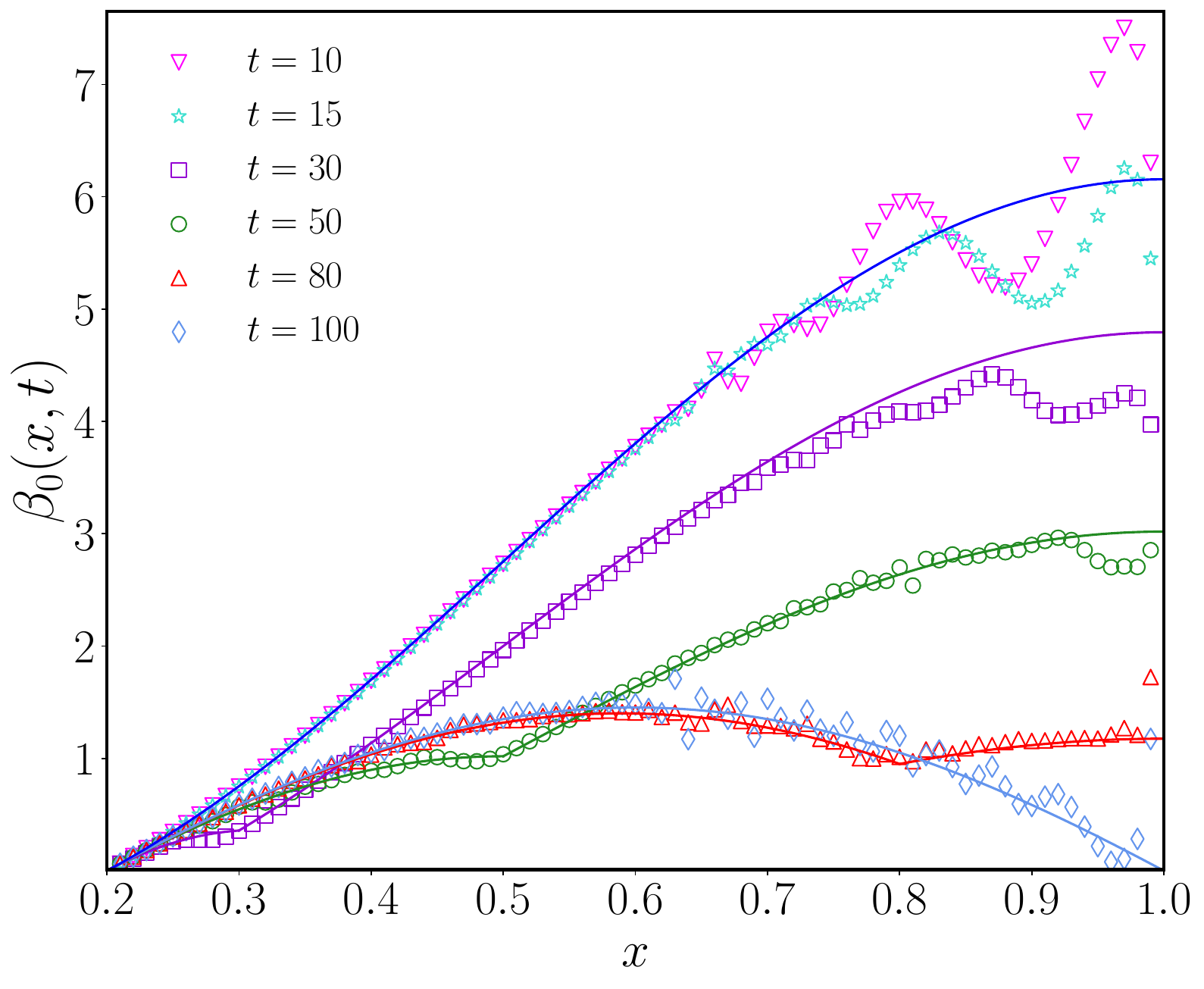}}
    \hspace{0.5cm}
\subfloat{\includegraphics[width=0.47\textwidth]{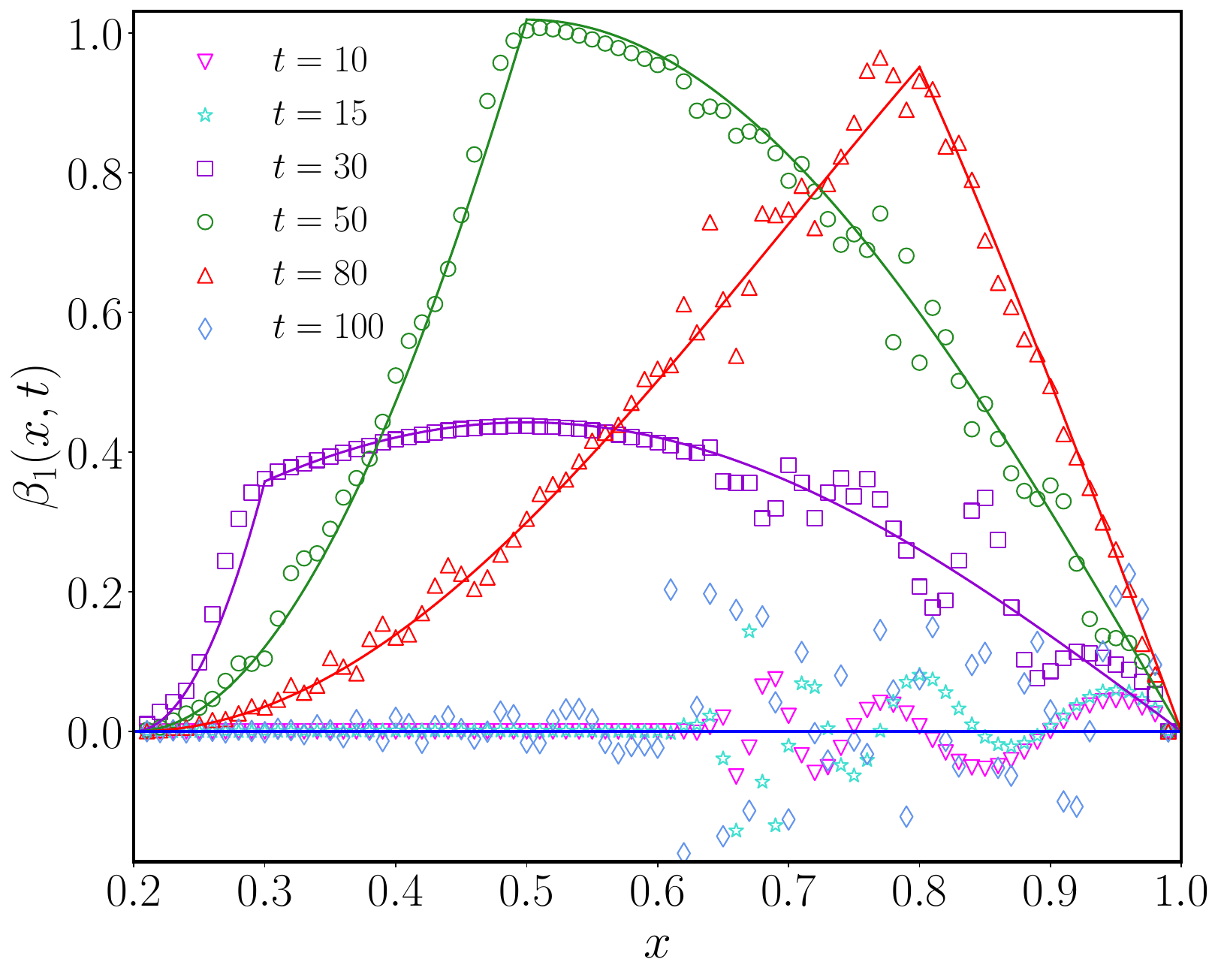}}
   \caption{Spatial profile of the weights $\beta_0(x,t)$ (left) and $\beta_1(x,t)$ (right) as function of $x=j/L$, for $L=100$ and a subsystem starting at site $d+1=21$. The symbols represent the numerical data obtained from \eqref{eq:CLhf} with $a=1/L$, and the solid lines are the corresponding CFT predictions \eqref{eq:beta01} with \eqref{eq:betadec}. }
     \label{fig:betaCLx020}
\end{figure}

In Fig.~\ref{fig:betaCLx020}, we show analogous plots for the decentered case, choosing $d = 20$ and $L=100$, such that $x_0=d/L= 0.2$. As discussed after Eq.~\eqref{eq:betadec}, for times $t < d$ the front has not yet reached the subsystem, thus $\beta_0(x,t)$ should correspond to the ground-state inverse temperature, while $\beta_1(x,t)=0$. This is indeed compatible with the numerical data, again up to oscillations close to the open boundary of the chain. For $t>d$ the weights become time dependent, and their behaviour changes at $j=t$. Although the curves are qualitatively similar to the $x_0=0$ case, the scaled weights $\beta_0(x,t)$ and $\beta_1(x,t)$ now do not collapse to a single curve in the regime $x<t$ and $x>t$, respectively. The agreement between the data and the CFT prediction is remarkably good for all times shown.

%
%

\section{Discussion}
\label{sec:discussion}

We studied the EH of free fermions after a local quench, where two half-chains of equal lengths are prepared in their ground state and subsequently joined together. In the CFT framework, we obtained the expression of the EH as a sum of weighted integrals with the holomorphic and anti-holomorphic components of the energy-momentum tensor. The corresponding inverse temperatures $\beta(x,t)$ and $\bar{\beta}(x,t)$ have been derived via suitable conformal mappings. They are described by functions that have extra zeros at $x=\pm t$ (additionally to the entangling point $x=x_0$), associated to the front created by the initial perturbation and propagating ballistically from the center. For the right-moving component $\beta(x,t)$, the additional zero at $x=t$ lies within the subsystem and thus implies an enhanced contribution to the entanglement, as compared to the left-moving part $\bar\beta(x,t)$. We verified that the entanglement entropy calculated via the inverse temperatures reproduces the known CFT result of \cite{SD-11}. Finally, we studied the local quench on the hopping chain numerically, introducing a suitable continuum limit of the lattice EH. This allowed us to compare the results directly to the CFT prediction, finding a good agreement.

It is important to stress that the continuum limit procedure is absolutely necessary to recover the CFT results. Indeed, while the real elements $t_1(x)$ in the first diagonal clearly dominate the EH matrix, their shape still significantly deviates from the CFT expression of $\beta_0(x,t)$, and the convergence is visible only after including several terms $t_{2p+1}(x)$ with $p>0$ in the sum \eqref{eq:beta01}. This mechanism is even more striking for the imaginary matrix elements $s_{2p}(x)$, where the decay in $p$ seems to be much slower, and one essentially needs to consider the complete sum in order to converge towards the CFT result.

It would be interesting to generalize the calculations for a local quench away from half filling. In the CFT setting, the only expected change is a rescaling of time with the Fermi velocity $v_F=\sin(q_Fa)$. However, the analysis of the continuum limit becomes more complicated, since for generic fillings the special structure of the matrices $C(t)$ and $H(t)$ are lost. Indeed, the matrix elements then contain both real and imaginary parts, and there is no separation between even/odd diagonals, which makes the numerical analysis of the sums in \eqref{eq:multipliers} more cumbersome. Nevertheless, we expect that this can be worked out and the results should again be well described by CFT, up to eventual oscillations.

Finally, another open question is whether the lattice EH could be approximated by a simpler matrix without long-range hopping terms. In fact, for hopping chains in their ground states, one finds a number of examples where the EH commutes with a tridiagonal matrix, with entries that correspond precisely to the discretized form of the CFT weight function \cite{P-04,CNV-19,BBEPV-24,BCNPV-24,Eis-25}. Obviously, due to the presence of two different weight functions, the situation must be more complicated for the local quench, as one would need to add at least one more diagonal. Identifying a faithful local approximation of the lattice EH would also be crucial to
study the local quench for interacting models. Indeed, in this context one needs a local ansatz for the EH as an input, which can be verified against numerical simulations \cite{DVZ-18,GMSCD-18}, and also forms the basis of learning the EH in experimental setups \cite{Kokailetal21a,Joshietal23}.




\section{Acknowledgements}

We thank Luca Capizzi, Jean-Marie Stéphan and Erik Tonni for fruitful discussions and correspondence.
This research was funded in whole by the Austrian Science Fund (FWF) Grant-DOI: 10.55776/P35434 and 10.55776/PAT3563424.

\appendix

\section{CFT calculations for the decentered case}
\label{app:A}

In this appendix, we show how the analytic continuation of Eq.~\eqref{eq:wprime} leads to the expressions in Eq.~\eqref{eq:betadec}, corresponding to different parameter regimes. A key step in this procedure is the resolution of the sign ambiguity in the square root,
$\xi(z) = \ir \sqrt{f(z)} = \pm \sqrt{-f(z)},$
which affects the final result after taking the analytic continuation $\tau \rightarrow \ir t$. To fix this ambiguity, we choose the sign of $\xi(z)$ by requiring that, after the analytic continuation, both $\xi(z)$ and $\xi_0$ lie in the upper half-plane (UHP), while $\bar{\xi}_0 = \mp \sqrt{-\bar{f}(z_0)}$ lies in the lower half-plane (LHP).

We begin by considering the regime $t < x_0 < x $, which describes the situation in which the propagating front has not yet entered the subsystem. In this case, we have
$\xi(z) = \sqrt{-f(z)}$ and $\xi_0 = \sqrt{-f(z_0)}$, so that both quantities lie in the UHP after analytic continuation. Similarly, we take $\bar{\xi}_0 = \sqrt{-\bar{f}(z_0)}$ so that it lies in the LHP. The same choice of signs applies to the terms in $\bar{\beta}(z)$, so we illustrate the computation only for $\beta(z)$, as follows
    \begin{equation}
       \label{eq:betatx0x}\beta(z)=\frac{\left[\sqrt{\frac{\sin \frac{\pi}{2L}(z+\ir \lambda)}{\sin \frac{\pi}{2L}(z-\ir \lambda)}}- \left( \sqrt{\frac{\sin \frac{\pi}{2L}( x_0+\ir \tau+\ir \lambda)}{\sin \frac{\pi}{2L}( x_0+\ir \tau-\ir \lambda)}}\right)\right]\left[\sqrt{\frac{\sin \frac{\pi}{2L}(z+\ir \lambda)}{\sin \frac{\pi}{2L}(z-\ir \lambda)}}- \left( \sqrt{\frac{\sin \frac{\pi}{2L}( x_0-\ir \tau-\ir \lambda)}{\sin \frac{\pi}{2L}( x_0-\ir \tau+\ir \lambda)}}\right)\right]}{\left[ \left( \sqrt{\frac{\sin \frac{\pi}{2L}( x_0+\ir \tau+\ir \lambda)}{\sin \frac{\pi}{2L}( x_0+\ir \tau-\ir \lambda)}}\right)-\left( \sqrt{\frac{\sin \frac{\pi}{2L}( x_0-\ir \tau-\ir \lambda)}{\sin \frac{\pi}{2L}( x_0-\ir \tau+\ir \lambda)}}\right)\right] \xi'(z)}, \quad t<x_0<x,
    \end{equation}
where $\xi'(z)$ is given by \eqref{eq:xiprime}.
We observe that each of the four terms in \eqref{eq:betatx0x} vanishes in the limit $\lambda = 0$, however, $\beta(z)$ remains finite. To show this, we rewrite the three terms involving square root differences by multiplying and dividing by their corresponding sums, obtaining 
\begin{equation}
\label{eq:ximxi0}
   (\xi(z)-\xi_{0})=\frac{-\sinh \left( \frac{\pi \lambda}{L}\right)\sin \frac{\pi(z-\ir \tau-x_0)}{2L} }{\sqrt{\sin\frac{\pi(\ir \lambda-z)}{2L} \sin\frac{\pi(\ir \lambda-\ir \tau -x_0)}{2L} } \left(\sqrt{\sin\frac{\pi(\ir \lambda+z)}{2L} \sin\frac{\pi(\ir \lambda-\ir \tau -x_0)}{2L} }+\sqrt{\sin\frac{\pi(\ir \lambda-z) }{2L}\sin\frac{\pi(\ir \lambda+\ir \tau +x_0)}{2L} }\right)},
\end{equation}
\begin{equation}
\label{eq:ximxi0bar}
    (\xi(z)-\Bar{\xi}_{0})=  \frac{\sinh\frac{\pi \lambda}{L} \sin\frac{\pi(z-\ir \tau+x_0)}{2L}}{\sqrt{\sin\frac{\pi(\ir \lambda-z)}{2L} \sin\frac{\pi(\ir \lambda-\ir \tau+x_0)}{2L} }\left(\sqrt{\sin\frac{\pi(\ir \lambda+z)}{2L} \sin\frac{\pi(\ir \lambda-\ir \tau+x_0)}{2L} }+\sqrt{\sin\frac{\pi(\ir \lambda-z)}{2L} \sin\frac{\pi(\ir \lambda+\ir \tau-x_0)}{2L} }\right)},
\end{equation}
\begin{equation}
\label{eq:xi0mxi0bar}
 (\xi_{0}-\Bar{\xi}_{0})=  \frac{\sinh\frac{\pi \lambda}{L} \sin\frac{\pi x_0}{L}}{\sqrt{\sin\frac{\pi(\ir \lambda-\ir \tau-x_0)}{2L} \sin\frac{\pi(\ir \lambda-\ir \tau+x_0)}{2L} }(\sqrt{\sin\frac{\pi(\ir \lambda+x_0+\ir \tau)}{2L} \sin\frac{\pi(\ir \lambda-\ir \tau+x_0)}{2L} }+\sqrt{\sin\frac{\pi(\ir \lambda-x0-\ir \tau)}{2L} \sin\frac{\pi(\ir \lambda+\ir \tau-x_0)}{2L} })}.
\end{equation}
In this way, the factors of $\sinh(\pi\lambda/L)$ cancel, and after simplification we obtain the following expression for $\beta(x,t)$ in the limit $\lambda = 0$, 
\begin{equation}
\label{eq:betaeqCFT}
\begin{split}
    \beta(x,t)&=\frac{L}{ \pi}\frac{\left(\cos \frac{\pi x_0}{L}-\cos \frac{\pi x}{L}\right)\left(\cosh \frac{\pi \tau}{L}-\cos \frac{\pi x_0}{L}\right)}{2\sin \frac{\pi x_0}{L}\sin\frac{\pi(x_0-\ir \tau)}{2L}\sin\frac{\pi(x_0+\ir \tau)}{2L}}=\frac{L}{ \pi} \frac{\left(\cos \frac{\pi x_0}{L}-\cos \frac{\pi x}{L}\right)}{\sin \frac{\pi x_0}{L}},
\end{split}
\end{equation}
which is time-independent. An identical expression is found for $\bar{\beta}(x,t)$.

We then move to the regime $ x_0 < x < t <L$, which corresponds to points in the subsystem located to the left of the propagating front, after it has entered the subsystem. To ensure that $\xi(z)$ remains in the UHP after analytic continuation, we must choose $\xi(z) = -\sqrt{-f(z)}$ and $\xi_0 = -\sqrt{-f(z_0)}$, while for $\bar{\xi}_0$ we keep the same sign of the previous case.  Again, the same sign conventions apply to both $\beta(z)$ and $\bar{\beta}(z)$, and we show only the former, 
\begin{equation}
\label{eq:x0xt}
\beta(z)=    \frac{\left[-\sqrt{\frac{\sin \frac{\pi}{2L}(z+\ir \lambda)}{\sin \frac{\pi}{2L}(z-\ir \lambda)}}- \left( -\sqrt{\frac{\sin \frac{\pi}{2L}( x_0+\ir \tau+\ir \lambda)}{\sin \frac{\pi}{2L}( x_0+\ir \tau-\ir \lambda)}}\right)\right]\left[-\sqrt{\frac{\sin \frac{\pi}{2L}(z+\ir \lambda)}{\sin \frac{\pi}{2L}(z-\ir \lambda)}}- \left( \sqrt{\frac{\sin \frac{\pi}{2L}( x_0-\ir \tau-\ir \lambda)}{\sin \frac{\pi}{2L}( x_0-\ir \tau+\ir \lambda)}}\right)\right]}{\left[ \left( -\sqrt{\frac{\sin \frac{\pi}{2L}( x_0+\ir \tau+\ir \lambda)}{\sin \frac{\pi}{2L}( x_0+\ir \tau-\ir \lambda)}}\right)-\left( \sqrt{\frac{\sin \frac{\pi}{2L}( x_0-\ir \tau-\ir \lambda)}{\sin \frac{\pi}{2L}( x_0-\ir \tau+\ir \lambda)}}\right)\right] \xi'(z)}, \quad x_0<x<t.
    \end{equation}
In this case, there are only two terms giving zeroes for $\lambda=0$, namely $\xi'(z)$ in the denominator, and the factor $(\xi(z)-\xi_0)$, that we thus rewrite as \eqref{eq:ximxi0}.
The calculation above, after setting $z=x+\ir \tau$ and $\lambda=0$, gives
\begin{equation}
   \beta(x,\tau) =\frac{2L}{\pi}\sin \frac{\pi}{2L}(x-x_0) \sqrt{\frac{1-\cos \frac{\pi(x+\ir \tau)}{L}}{\cos \frac{\pi x_0}{L}-\cosh \frac{\pi \tau}{L}}}\sqrt{ \frac{\sin \frac{\pi(\ir \tau-x_0)}{2L}}{\sin \frac{\pi(x_0+\ir \tau)}{2L}}}
\end{equation}
and after analytic continuation $\tau \rightarrow \ir t$,
\begin{equation}
\label{eq:betax0xt}
\beta(x,t)=\frac{2L}{\pi}\sin \left[ \frac{\pi}{2L}(x-x_0)\right] \frac{\sin \frac{\pi}{2L}(t-x)}{\sin \frac{\pi}{2L}(t-x_0)},
\end{equation}
which is the first term in the expression of $\beta_0(x,t)$ given in \eqref{eq:betadec}. Analogously from $\bar{\beta}(z)$, with the same choice of the signs as in \eqref{eq:x0xt}, after analytic continuation we get
\begin{equation}
\label{eq:betabarx0xt}
\bar{\beta}(x,t)=\frac{2L}{\pi}\sin \left[ \frac{\pi}{2L}(x-x_0)\right] \frac{\sin \frac{\pi}{2L}(x+t)}{\sin \frac{\pi}{2L}(x_0+t)}.
\end{equation}
Then, we consider the regime $ x_0 < t < x < L $ , which describes the points of the subsystem located to the right of the front, before it reaches the right boundary.
In this case, the term $\bar{\beta}(z)$ keeps the same sign conventions as in equation \eqref{eq:x0xt}, and therefore corresponds to expression \eqref{eq:betabarx0xt}. Meanwhile, in the term $\beta(z)$, we have $\xi(z) = \sqrt{-f(z)}$, as in the first case since $ t < x $. However, $\xi_0$ and $\bar{\xi}_0$ follow the same sign convention as in the second case, where $ t < x_0 $. Thus, we have
\begin{equation}
\label{eq:betax0tx}
\beta(z)=\frac{\left[\sqrt{\frac{\sin \frac{\pi}{2L}(z+\ir \lambda)}{\sin \frac{\pi}{2L}(z-\ir \lambda)}}- \left( -\sqrt{\frac{\sin \frac{\pi}{2L}( x_0+\ir \tau+\ir \lambda)}{\sin \frac{\pi}{2L}( x_0+\ir \tau-\ir \lambda)}}\right)\right]\left[\sqrt{\frac{\sin \frac{\pi}{2L}(z+\ir \lambda)}{\sin \frac{\pi}{2L}(z-\ir \lambda)}}- \left( \sqrt{\frac{\sin \frac{\pi}{2L}( x_0-\ir \tau-\ir \lambda)}{\sin \frac{\pi}{2L}( x_0-\ir \tau+\ir \lambda)}}\right)\right]}{\left[ \left( -\sqrt{\frac{\sin \frac{\pi}{2L}( x_0+\ir \tau+\ir \lambda)}{\sin \frac{\pi}{2L}( x_0+\ir \tau-\ir \lambda)}}\right)-\left( \sqrt{\frac{\sin \frac{\pi}{2L}( x_0-\ir \tau-\ir \lambda)}{\sin \frac{\pi}{2L}( x_0-\ir \tau+\ir \lambda)}}\right)\right] \xi'(z)}.
\end{equation}
In this case, to extract the result in the limit $\lambda = 0$, we need to rewrite the term $(\xi(z) - \bar{\xi}_0)$ as in \eqref{eq:ximxi0bar}, since it vanishes in this limit along with $\xi'(z)$. In contrast, the other terms involve square roots that are added with the same sign and thus remain finite.
After performing the calculations and analytical continuation, we get
\begin{equation}
\beta(x,t)=\frac{2L}{\pi}\frac{\sin \frac{\pi}{2L}(x+x_0)\sin \frac{\pi}{2L}(x-t)}{\sin\frac{\pi}{2L}(t+x_0)}.
\end{equation}
After the front gets reflected at the right boundary,  the terms $\beta(z)$ and $\bar{\beta}(z)$ switch their sign conventions. Specifically, for the points to located to the right of the front, $\beta(z)$  follows the same sign choice as in equation \eqref{eq:x0xt}, resulting in expression \eqref{eq:betax0xt}, while $\bar{\beta}(z)$ adopts the sign convention of \eqref{eq:betax0tx}.
%
%
%
For the point located between the left boundary $x_0$ and the front, $\beta(z)$ and $\bar{\beta}(z)$ are again given by \eqref{eq:betax0xt} and \eqref{eq:betabarx0xt}, respectively.

Finally, after the front reaches again the left boundary at time $t=2L-x_0$ and exits the subsystem, the $\beta(z)$ and $\bar{\beta}(z)$ are again given by \eqref{eq:betaeqCFT}.

\section{Calculation of the continuum limit}
\label{app:CL}

In this appendix, we illustrate the derivation of the continuum limit of the EH after a local quench. We start from expression \eqref{eq:EHCL}, that we report also here for convenience,
\begin{equation}
    \mathcal{H}=\sum_j t_0(j)c_j^{\dagger}c_j+\sum_j\sum_{r\geq 1}\left[ t_r(j+r/2)  \left( c_j^{\dagger}c_{j+r}+c_{j+r}^{\dagger}c_j\right)+ \ir s_r(j+r/2)\left( c_j^{\dagger }c_{j+r}-c_{j+r}^{\dagger}c_j\right)\right].
\end{equation}
While the real hopping terms are the same as in the equilibrium case \cite{ETP-19}, the imaginary terms represent new contributions specific to the non-equilibrium setting. Therefore, our main focus will be on their treatment. As first step, we replace the lattice operators with the right/left-moving fields
     \begin{equation}
    \begin{split}
        c_j \quad \rightarrow  \quad & \sqrt{a} \left(\eE^{\ir k_Fx}\psi(x)+\eE^{-\ir k_Fx}\bar{\psi}(x) \right),\\
         c_{j+r} \quad \rightarrow  \quad & \sqrt{a} \left(\eE^{\ir k_F(x+ra)}\psi(x+ra)+\eE^{-\ir k_F(x+ra)}\bar{\psi}(x+ra) \right).
    \end{split}
\end{equation}
The expression of the hopping terms after this substitution is,
\begin{equation}
   \begin{split}
     c_{j}^{\dagger}c_{j+r}+ c_{j+r}^{\dagger}c_j \quad \rightarrow \quad  a \cos(r a k_F)&[\psi^{\dagger}(x)\psi(x+r a)+\bar{\psi}^{\dagger}(x)\bar{\psi}(x+r a)+ h.c.]\\
        + a\sin(r a k_F)&[\ir (\psi^{\dagger}(x)\psi(x+r a)-\bar{\psi}^{\dagger}(x)\bar{\psi}(x+r a))+h.c.]
   \end{split}
\end{equation}
and
\begin{equation}
   \begin{split}
c_{j}^{\dagger}c_{j+r}-c_{j+r}^{\dagger}c_j \quad \rightarrow \quad  a \cos(r a k_F)&[\psi^{\dagger}(x)\psi(x+r a)+\bar{\psi}^{\dagger}(x)\bar{\psi}(x+r a) -h.c.]\\
        + a\sin(r a k_F)&[\ir (\psi^{\dagger}(x)\psi(x+r a)-\bar{\psi}^{\dagger}(x)\bar{\psi}(x+r a))-h.c.].
   \end{split}
\end{equation}
We now proceed by expanding all the terms as $a\rightarrow 0$, which  gives for the hopping amplitudes $t(j+r/2)\rightarrow t(x)+ a r t'(x)/2+\mathcal{O}(a^2)$ and $s(j+r/2)\rightarrow s(x)+ a r s'(x)/2+\mathcal{O}(a^2)$, while the on-site amplitude goes to $t_0(j)\rightarrow t(x)$.
Moreover, we expand the fields as 
\begin{equation}
    \begin{split}
        \psi(x+r a)\sim [\psi(x)+r a\partial_x\psi(x)],\qquad  \bar{\psi}(x+r a)\sim [\bar{\psi}(x)+r a\partial_x\bar{\psi}(x)]
    \end{split}
\end{equation}
and the same for their hermitian conjugates. With these substitutions, the EH after taking the continuum limit becomes
\begin{equation}
\label{eq:EHafterCL}
    \begin{split}
        \mathcal{H}&= \int_A\dd x t_0(x)[\psi^{\dagger}\psi+\bar{\psi}^{\dagger}\bar{\psi}]+\\
        &\int_A \dd x\sum_{r \geq 0} \left[ t_r(x)+\frac{r a}{2}t_r'(x)\right] \cos(r a k_F)[2(\psi^{\dagger}\psi+\bar{\psi}^{\dagger}\bar{\psi} )+r a \partial_x(\psi^{\dagger}\psi+\bar{\psi}^{\dagger}\bar{\psi})]\\
        &\int_A \dd x \sum_{r \geq 0} \left[ t_r(x)+\frac{r a}{2}t_r'(x)\right]\sin(r ak_F) \ir r a [\psi^{\dagger}\psi'-\bar{\psi}^{\dagger}\bar{\psi}'-\psi^{\dagger\prime}\psi+\bar{\psi}^{\dagger\prime}\bar{\psi}]\\
        &\int_A \dd x \sum_{r \geq 0} \ir\left[ s_r(x)+\frac{r a}{2}s_r'(x)\right] \cos(r a k_F) ra [ (\psi^{\dagger}\psi'+\bar{\psi}^{\dagger}\bar{\psi}'-\psi^{\dagger\prime}\psi-\bar{\psi}^{\dagger\prime}\bar{\psi})]\\
        &-\int_A \dd x\sum_{r \geq 0} \left[ s_r(x)+\frac{r a}{2}s_r'(x)\right] \sin(r ak_F) [2(\psi^{\dagger}\psi-\bar{\psi}^{\dagger}\bar{\psi})+ra(\psi^{\dagger}\psi'+\psi^{\dagger\prime}\psi-\bar{\psi}^{\dagger}\bar{\psi}'-\bar{\psi}^{\dagger\prime}\bar{\psi}) ].
    \end{split}
\end{equation}
This expression can be further worked out by neglecting terms $\mathcal{O}(a^2)$, and integrating by parts. As illustrated in \cite{ETP-19}, the hopping terms after integrating by parts the second line of \eqref{eq:EHafterCL} give
\begin{equation}
    \begin{split}
        & \int_A \dd x \left[ t_0(x)+2 \sum_{r\geq 0} t_r(x)\cos(rak_F)\right](\psi^{\dagger}\psi+\bar{\psi}^{\dagger}\bar{\psi})\\
            &- \int_A \dd x \sum_{r \geq 0} t_r(x) 2 a r \sin(a r k_F) \frac{1}{2}\left(\psi^{\dagger}(-\ir \partial_x)\psi -\bar{\psi}^{\dagger}(-\ir \partial_x)\bar{\psi}+ h.c.\right)
    \end{split}
\end{equation}
We now consider the continuum limit of the remaining terms. The fourth term in \eqref{eq:EHafterCL}, after neglecting higher order terms in the lattice parameter, can be rewritten as 
\begin{equation}
     -\int_A \dd x 2 a\sum_{r \geq 0} rs_r(x)   \cos(a r k_F) \frac{1}{2}\left(\psi^{\dagger}(-\ir \partial_x)\psi -\bar{\psi}^{\dagger}(-\ir \partial_x)\bar{\psi}+ h.c.\right)
\end{equation}
Finally, we consider the the last term of \eqref{eq:EHafterCL}. Expanding all the terms and neglecting those of order $\mathcal{O}(a^2)$, we get
\begin{equation}
   \begin{split}
      & - \int_A \dd x 2 \sum_{r \geq 0} s_r(x)\sin(rak_F)[\psi^{\dagger}(x)\psi(x)-\bar{\psi}^{\dagger}(x)\bar{\psi}(x)]+\\
        - &\int_A \dd x 2 \sum_{r \geq 0}s'_r(x)\frac{r a}{2}\sin(rak_F)[\psi^{\dagger}(x)\psi(x)-\bar{\psi}^{\dagger}(x)\bar{\psi}(x)]+\\
         - &\int_A \dd x  \sum_{r \geq 0}s_r(x)r a\sin(rak_F)\partial_x(\psi^{\dagger}\psi-\bar{\psi}^{\dagger}\bar{\psi}).
   \end{split}
\end{equation}
We can verify that the last two lines cancel integrating by parts, and we are left with the first line.
 So putting all the terms together, we have
    \begin{equation}
       \begin{split}
            \mathcal{H}=&  \int_A \dd x \left[ t_0(x)+2 \sum_{r\geq 0} t_r(x)\cos(rak_F)\right](\psi^{\dagger}\psi+\bar{\psi}^{\dagger}\bar{\psi})\\
            &- \int_A \dd x \sum_{r \geq 0} t_r(x) 2 a r \sin(a r k_F) \frac{1}{2}\left(\psi^{\dagger}(-\ir \partial_x)\psi -\bar{\psi}^{\dagger}(-\ir \partial_x)\bar{\psi}+ h.c.\right)\\
             & - \int_A \dd x 2 a\sum_{r \geq 0} rs_r(x)   \cos(a r k_F)\frac{1}{2}\left(\psi^{\dagger}(-\ir \partial_x)\psi +\bar{\psi}^{\dagger}(-\ir \partial_x)\bar{\psi}+ h.c.\right) \\&- \int_A \dd x  2\sum_{r \geq 0} s_r(x)\sin(rak_F)[\psi^{\dagger}\psi-\bar{\psi}^{\dagger}\bar{\psi}],
       \end{split}
    \end{equation}
that, after using the definitions in \eqref{eq:current}, \eqref{eq:TTensor}, and \eqref{eq:multipliers}, gives exactly the result \eqref{eq:EHcontinuumlimit}.


\begin{thebibliography}{64}%
\makeatletter
\providecommand \@ifxundefined [1]{%
 \@ifx{#1\undefined}
}%
\providecommand \@ifnum [1]{%
 \ifnum #1\expandafter \@firstoftwo
 \else \expandafter \@secondoftwo
 \fi
}%
\providecommand \@ifx [1]{%
 \ifx #1\expandafter \@firstoftwo
 \else \expandafter \@secondoftwo
 \fi
}%
\providecommand \natexlab [1]{#1}%
\providecommand \enquote  [1]{``#1''}%
\providecommand \bibnamefont  [1]{#1}%
\providecommand \bibfnamefont [1]{#1}%
\providecommand \citenamefont [1]{#1}%
\providecommand \href@noop [0]{\@secondoftwo}%
\providecommand \href [0]{\begingroup \@sanitize@url \@href}%
\providecommand \@href[1]{\@@startlink{#1}\@@href}%
\providecommand \@@href[1]{\endgroup#1\@@endlink}%
\providecommand \@sanitize@url [0]{\catcode `\\12\catcode `\$12\catcode `\&12\catcode `\#12\catcode `\^12\catcode `\_12\catcode `\%12\relax}%
\providecommand \@@startlink[1]{}%
\providecommand \@@endlink[0]{}%
\providecommand \url  [0]{\begingroup\@sanitize@url \@url }%
\providecommand \@url [1]{\endgroup\@href {#1}{\urlprefix }}%
\providecommand \urlprefix  [0]{URL }%
\providecommand \Eprint [0]{\href }%
\providecommand \doibase [0]{https://doi.org/}%
\providecommand \selectlanguage [0]{\@gobble}%
\providecommand \bibinfo  [0]{\@secondoftwo}%
\providecommand \bibfield  [0]{\@secondoftwo}%
\providecommand \translation [1]{[#1]}%
\providecommand \BibitemOpen [0]{}%
\providecommand \bibitemStop [0]{}%
\providecommand \bibitemNoStop [0]{.\EOS\space}%
\providecommand \EOS [0]{\spacefactor3000\relax}%
\providecommand \BibitemShut  [1]{\csname bibitem#1\endcsname}%
\let\auto@bib@innerbib\@empty
\bibitem [{\citenamefont {Amico}\ \emph {et~al.}(2008)\citenamefont {Amico}, \citenamefont {Fazio}, \citenamefont {Osterloh},\ and\ \citenamefont {Vedral}}]{AFOV-08}%
  \BibitemOpen
  \bibfield  {author} {\bibinfo {author} {\bibfnamefont {L.}~\bibnamefont {Amico}}, \bibinfo {author} {\bibfnamefont {R.}~\bibnamefont {Fazio}}, \bibinfo {author} {\bibfnamefont {A.}~\bibnamefont {Osterloh}},\ and\ \bibinfo {author} {\bibfnamefont {V.}~\bibnamefont {Vedral}},\ }\bibfield  {title} {\bibinfo {title} {Entanglement in many-body systems},\ }\href {https://doi.org/10.1103/RevModPhys.80.517} {\bibfield  {journal} {\bibinfo  {journal} {Rev. Mod. Phys.}\ }\textbf {\bibinfo {volume} {80}},\ \bibinfo {pages} {517} (\bibinfo {year} {2008})}\BibitemShut {NoStop}%
\bibitem [{\citenamefont {Calabrese}\ \emph {et~al.}(2009)\citenamefont {Calabrese}, \citenamefont {Cardy},\ and\ \citenamefont {Doyon}}]{CCD-09}%
  \BibitemOpen
  \bibfield  {author} {\bibinfo {author} {\bibfnamefont {P.}~\bibnamefont {Calabrese}}, \bibinfo {author} {\bibfnamefont {J.}~\bibnamefont {Cardy}},\ and\ \bibinfo {author} {\bibfnamefont {B.}~\bibnamefont {Doyon}},\ }\bibfield  {title} {\bibinfo {title} {Entanglement entropy in extended quantum systems},\ }\href {https://doi.org/10.1088/1751-8121/42/50/500301} {\bibfield  {journal} {\bibinfo  {journal} {J. Phys. A: Math. Theor.}\ }\textbf {\bibinfo {volume} {42}},\ \bibinfo {pages} {500301} (\bibinfo {year} {2009})}\BibitemShut {NoStop}%
\bibitem [{\citenamefont {Laflorencie}(2016)}]{Laflorencie-16}%
  \BibitemOpen
  \bibfield  {author} {\bibinfo {author} {\bibfnamefont {N.}~\bibnamefont {Laflorencie}},\ }\bibfield  {title} {\bibinfo {title} {Quantum entanglement in condensed matter systems},\ }\href {https://doi.org/https://doi.org/10.1016/j.physrep.2016.06.008} {\bibfield  {journal} {\bibinfo  {journal} {Phys. Reports}\ }\textbf {\bibinfo {volume} {646}},\ \bibinfo {pages} {1} (\bibinfo {year} {2016})}\BibitemShut {NoStop}%
\bibitem [{\citenamefont {Vidal}\ \emph {et~al.}(2003)\citenamefont {Vidal}, \citenamefont {Latorre}, \citenamefont {Rico},\ and\ \citenamefont {Kitaev}}]{VLRK-03}%
  \BibitemOpen
  \bibfield  {author} {\bibinfo {author} {\bibfnamefont {G.}~\bibnamefont {Vidal}}, \bibinfo {author} {\bibfnamefont {J.~I.}\ \bibnamefont {Latorre}}, \bibinfo {author} {\bibfnamefont {E.}~\bibnamefont {Rico}},\ and\ \bibinfo {author} {\bibfnamefont {A.}~\bibnamefont {Kitaev}},\ }\bibfield  {title} {\bibinfo {title} {Entanglement in quantum critical phenomena},\ }\href {https://doi.org/10.1103/PhysRevLett.90.227902} {\bibfield  {journal} {\bibinfo  {journal} {Phys. Rev. Lett.}\ }\textbf {\bibinfo {volume} {90}},\ \bibinfo {pages} {227902} (\bibinfo {year} {2003})}\BibitemShut {NoStop}%
\bibitem [{\citenamefont {Calabrese}\ and\ \citenamefont {Cardy}(2009)}]{CC-09}%
  \BibitemOpen
  \bibfield  {author} {\bibinfo {author} {\bibfnamefont {P.}~\bibnamefont {Calabrese}}\ and\ \bibinfo {author} {\bibfnamefont {J.}~\bibnamefont {Cardy}},\ }\bibfield  {title} {\bibinfo {title} {Entanglement entropy and conformal field theory},\ }\href {https://doi.org/10.1088/1751-8113/42/50/504005} {\bibfield  {journal} {\bibinfo  {journal} {J. Phys. A: Math. Theor.}\ }\textbf {\bibinfo {volume} {42}},\ \bibinfo {pages} {504005} (\bibinfo {year} {2009})}\BibitemShut {NoStop}%
\bibitem [{\citenamefont {Eisert}\ \emph {et~al.}(2010)\citenamefont {Eisert}, \citenamefont {Cramer},\ and\ \citenamefont {Plenio}}]{ECP-10}%
  \BibitemOpen
  \bibfield  {author} {\bibinfo {author} {\bibfnamefont {J.}~\bibnamefont {Eisert}}, \bibinfo {author} {\bibfnamefont {M.}~\bibnamefont {Cramer}},\ and\ \bibinfo {author} {\bibfnamefont {M.~B.}\ \bibnamefont {Plenio}},\ }\bibfield  {title} {\bibinfo {title} {Colloquium: Area laws for the entanglement entropy},\ }\href {https://doi.org/10.1103/RevModPhys.82.277} {\bibfield  {journal} {\bibinfo  {journal} {Rev. Mod. Phys.}\ }\textbf {\bibinfo {volume} {82}},\ \bibinfo {pages} {277} (\bibinfo {year} {2010})}\BibitemShut {NoStop}%
\bibitem [{\citenamefont {Calabrese}\ and\ \citenamefont {Cardy}(2016)}]{CC-16}%
  \BibitemOpen
  \bibfield  {author} {\bibinfo {author} {\bibfnamefont {P.}~\bibnamefont {Calabrese}}\ and\ \bibinfo {author} {\bibfnamefont {J.}~\bibnamefont {Cardy}},\ }\bibfield  {title} {\bibinfo {title} {Quantum quenches in 1+1 dimensional conformal field theories},\ }\href {https://doi.org/10.1088/1742-5468/2016/06/064003} {\bibfield  {journal} {\bibinfo  {journal} {J. Stat. Mech.}\ }\textbf {\bibinfo {volume} {2016}},\ \bibinfo {pages} {064003} (\bibinfo {year} {2016})}\BibitemShut {NoStop}%
\bibitem [{\citenamefont {Calabrese}(2018)}]{Calabrese-18}%
  \BibitemOpen
  \bibfield  {author} {\bibinfo {author} {\bibfnamefont {P.}~\bibnamefont {Calabrese}},\ }\bibfield  {title} {\bibinfo {title} {Entanglement and thermodynamics in non-equilibrium isolated quantum systems},\ }\href {https://doi.org/10.1016/j.physa.2017.10.011} {\bibfield  {journal} {\bibinfo  {journal} {Physica A: Stat. Mech. Appl.}\ }\textbf {\bibinfo {volume} {504}},\ \bibinfo {pages} {31} (\bibinfo {year} {2018})}\BibitemShut {NoStop}%
\bibitem [{\citenamefont {Dalmonte}\ \emph {et~al.}(2022)\citenamefont {Dalmonte}, \citenamefont {Eisler}, \citenamefont {Falconi},\ and\ \citenamefont {Vermersch}}]{DEFV-23}%
  \BibitemOpen
  \bibfield  {author} {\bibinfo {author} {\bibfnamefont {M.}~\bibnamefont {Dalmonte}}, \bibinfo {author} {\bibfnamefont {V.}~\bibnamefont {Eisler}}, \bibinfo {author} {\bibfnamefont {M.}~\bibnamefont {Falconi}},\ and\ \bibinfo {author} {\bibfnamefont {B.}~\bibnamefont {Vermersch}},\ }\bibfield  {title} {\bibinfo {title} {Entanglement {Hamiltonians}: From field theory to lattice models and experiments},\ }\href {https://doi.org/https://doi.org/10.1002/andp.202200064} {\bibfield  {journal} {\bibinfo  {journal} {Ann. Phys.}\ }\textbf {\bibinfo {volume} {534}},\ \bibinfo {pages} {2200064} (\bibinfo {year} {2022})}\BibitemShut {NoStop}%
\bibitem [{\citenamefont {Bisognano}\ and\ \citenamefont {Wichmann}(1975)}]{BW-75}%
  \BibitemOpen
  \bibfield  {author} {\bibinfo {author} {\bibfnamefont {J.~J.}\ \bibnamefont {Bisognano}}\ and\ \bibinfo {author} {\bibfnamefont {E.~H.}\ \bibnamefont {Wichmann}},\ }\bibfield  {title} {\bibinfo {title} {On the duality condition for a hermitian scalar field},\ }\href {https://doi.org/10.1063/1.522605} {\bibfield  {journal} {\bibinfo  {journal} {J. Math. Phys.}\ }\textbf {\bibinfo {volume} {16}},\ \bibinfo {pages} {985} (\bibinfo {year} {1975})}\BibitemShut {NoStop}%
\bibitem [{\citenamefont {Bisognano}\ and\ \citenamefont {Wichmann}(1976)}]{BW-76}%
  \BibitemOpen
  \bibfield  {author} {\bibinfo {author} {\bibfnamefont {J.~J.}\ \bibnamefont {Bisognano}}\ and\ \bibinfo {author} {\bibfnamefont {E.~H.}\ \bibnamefont {Wichmann}},\ }\bibfield  {title} {\bibinfo {title} {On the duality condition for quantum fields},\ }\href {https://doi.org/10.1063/1.522898} {\bibfield  {journal} {\bibinfo  {journal} {J. Math. Phys.}\ }\textbf {\bibinfo {volume} {17}},\ \bibinfo {pages} {303} (\bibinfo {year} {1976})}\BibitemShut {NoStop}%
\bibitem [{\citenamefont {Hislop}\ and\ \citenamefont {Longo}(1982)}]{HL-82}%
  \BibitemOpen
  \bibfield  {author} {\bibinfo {author} {\bibfnamefont {P.}~\bibnamefont {Hislop}}\ and\ \bibinfo {author} {\bibfnamefont {R.}~\bibnamefont {Longo}},\ }\bibfield  {title} {\bibinfo {title} {Modular structure of the local algebras associated with the free massless scalar field theory},\ }\href {https://doi.org/10.1007/BF01208372} {\bibfield  {journal} {\bibinfo  {journal} {Commun.Math. Phys.}\ }\textbf {\bibinfo {volume} {84}},\ \bibinfo {pages} {71–85} (\bibinfo {year} {1982})}\BibitemShut {NoStop}%
\bibitem [{\citenamefont {Casini}\ \emph {et~al.}(2011)\citenamefont {Casini}, \citenamefont {Huerta},\ and\ \citenamefont {Myers}}]{CHM-11}%
  \BibitemOpen
  \bibfield  {author} {\bibinfo {author} {\bibfnamefont {H.}~\bibnamefont {Casini}}, \bibinfo {author} {\bibfnamefont {M.}~\bibnamefont {Huerta}},\ and\ \bibinfo {author} {\bibfnamefont {R.}~\bibnamefont {Myers}},\ }\bibfield  {title} {\bibinfo {title} {Towards a derivation of holographic entanglement entropy},\ }\href {https://doi.org/10.1007/JHEP05(2011)036} {\bibfield  {journal} {\bibinfo  {journal} {J. High Energy Phys.}\ }\textbf {\bibinfo {volume} {2011}}}\BibitemShut {NoStop}%
\bibitem [{\citenamefont {Cardy}\ and\ \citenamefont {Tonni}(2016)}]{CT-16}%
  \BibitemOpen
  \bibfield  {author} {\bibinfo {author} {\bibfnamefont {J.}~\bibnamefont {Cardy}}\ and\ \bibinfo {author} {\bibfnamefont {E.}~\bibnamefont {Tonni}},\ }\bibfield  {title} {\bibinfo {title} {Entanglement {Hamiltonians} in two-dimensional conformal field theory},\ }\href {https://doi.org/10.1088/1742-5468/2016/12/123103} {\bibfield  {journal} {\bibinfo  {journal} {J. Stat. Mech.}\ }\textbf {\bibinfo {volume} {2016}},\ \bibinfo {pages} {123103} (\bibinfo {year} {2016})}\BibitemShut {NoStop}%
\bibitem [{\citenamefont {Wong}\ \emph {et~al.}(2013)\citenamefont {Wong}, \citenamefont {Klich}, \citenamefont {Zayas},\ and\ \citenamefont {Vaman}}]{WKPZV-13}%
  \BibitemOpen
  \bibfield  {author} {\bibinfo {author} {\bibfnamefont {G.}~\bibnamefont {Wong}}, \bibinfo {author} {\bibfnamefont {I.}~\bibnamefont {Klich}}, \bibinfo {author} {\bibfnamefont {L.~A.~P.}\ \bibnamefont {Zayas}},\ and\ \bibinfo {author} {\bibfnamefont {D.}~\bibnamefont {Vaman}},\ }\bibfield  {title} {\bibinfo {title} {Entanglement temperature and entanglement entropy of excited states},\ }\href {https://doi.org/10.1007/JHEP12(2013)020} {\bibfield  {journal} {\bibinfo  {journal} {J. High Energy Phys.}\ }\textbf {\bibinfo {volume} {2013}}\bibinfo  {number} { (12)},\ \bibinfo {pages} {20}}\BibitemShut {NoStop}%
\bibitem [{\citenamefont {Arias}\ \emph {et~al.}(2017)\citenamefont {Arias}, \citenamefont {Blanco}, \citenamefont {Casini},\ and\ \citenamefont {Huerta}}]{ABCH-17}%
  \BibitemOpen
\bibfield  {number} {  }\bibfield  {author} {\bibinfo {author} {\bibfnamefont {R.~E.}\ \bibnamefont {Arias}}, \bibinfo {author} {\bibfnamefont {D.~D.}\ \bibnamefont {Blanco}}, \bibinfo {author} {\bibfnamefont {H.}~\bibnamefont {Casini}},\ and\ \bibinfo {author} {\bibfnamefont {M.}~\bibnamefont {Huerta}},\ }\bibfield  {title} {\bibinfo {title} {Local temperatures and local terms in modular {Hamiltonians}},\ }\href {https://doi.org/10.1103/PhysRevD.95.065005} {\bibfield  {journal} {\bibinfo  {journal} {Phys. Rev. D}\ }\textbf {\bibinfo {volume} {95}},\ \bibinfo {pages} {065005} (\bibinfo {year} {2017})}\BibitemShut {NoStop}%
\bibitem [{\citenamefont {Casini}\ and\ \citenamefont {Huerta}(2009)}]{CH-09}%
  \BibitemOpen
  \bibfield  {author} {\bibinfo {author} {\bibfnamefont {H.}~\bibnamefont {Casini}}\ and\ \bibinfo {author} {\bibfnamefont {M.}~\bibnamefont {Huerta}},\ }\bibfield  {title} {\bibinfo {title} {Reduced density matrix and internal dynamics for multicomponent regions},\ }\href {https://doi.org/10.1088/0264-9381/26/18/185005} {\bibfield  {journal} {\bibinfo  {journal} {Class. and Quant Grav.}\ }\textbf {\bibinfo {volume} {26}},\ \bibinfo {pages} {185005} (\bibinfo {year} {2009})}\BibitemShut {NoStop}%
\bibitem [{\citenamefont {Longo}\ \emph {et~al.}(2010)\citenamefont {Longo}, \citenamefont {Martinetti},\ and\ \citenamefont {Rehren}}]{LMR-10}%
  \BibitemOpen
  \bibfield  {author} {\bibinfo {author} {\bibfnamefont {R.}~\bibnamefont {Longo}}, \bibinfo {author} {\bibfnamefont {P.}~\bibnamefont {Martinetti}},\ and\ \bibinfo {author} {\bibfnamefont {K.-H.}\ \bibnamefont {Rehren}},\ }\bibfield  {title} {\bibinfo {title} {Geometric modular action for disjoint intervals and boundary conformal field theory},\ }\href {https://doi.org/10.1142/S0129055X10003977} {\bibfield  {journal} {\bibinfo  {journal} {Rev. in Math. Phys.}\ }\textbf {\bibinfo {volume} {22}},\ \bibinfo {pages} {331} (\bibinfo {year} {2010})}\BibitemShut {NoStop}%
\bibitem [{\citenamefont {Arias}\ \emph {et~al.}(2018)\citenamefont {Arias}, \citenamefont {Casini}, \citenamefont {Huerta},\ and\ \citenamefont {Pontello}}]{ACHP-18}%
  \BibitemOpen
  \bibfield  {author} {\bibinfo {author} {\bibfnamefont {R.~E.}\ \bibnamefont {Arias}}, \bibinfo {author} {\bibfnamefont {H.}~\bibnamefont {Casini}}, \bibinfo {author} {\bibfnamefont {M.}~\bibnamefont {Huerta}},\ and\ \bibinfo {author} {\bibfnamefont {D.}~\bibnamefont {Pontello}},\ }\bibfield  {title} {\bibinfo {title} {Entropy and modular {Hamiltonian} for a free chiral scalar in two intervals},\ }\href {https://doi.org/10.1103/PhysRevD.98.125008} {\bibfield  {journal} {\bibinfo  {journal} {Phys. Rev. D}\ }\textbf {\bibinfo {volume} {98}},\ \bibinfo {pages} {125008} (\bibinfo {year} {2018})}\BibitemShut {NoStop}%
\bibitem [{\citenamefont {Mintchev}\ and\ \citenamefont {Tonni}(2021{\natexlab{a}})}]{MT-21a}%
  \BibitemOpen
  \bibfield  {author} {\bibinfo {author} {\bibfnamefont {M.}~\bibnamefont {Mintchev}}\ and\ \bibinfo {author} {\bibfnamefont {E.}~\bibnamefont {Tonni}},\ }\bibfield  {title} {\bibinfo {title} {Modular {H}amiltonians for the massless {D}irac field in the presence of a boundary},\ }\href {https://doi.org/10.1007/jhep03(2021)204} {\bibfield  {journal} {\bibinfo  {journal} {J. High Energy Phys.}\ }\textbf {\bibinfo {volume} {2021}}\bibinfo  {number} { (3)},\ \bibinfo {pages} {47}}\BibitemShut {NoStop}%
\bibitem [{\citenamefont {Mintchev}\ and\ \citenamefont {Tonni}(2021{\natexlab{b}})}]{MT-21b}%
  \BibitemOpen
\bibfield  {number} {  }\bibfield  {author} {\bibinfo {author} {\bibfnamefont {M.}~\bibnamefont {Mintchev}}\ and\ \bibinfo {author} {\bibfnamefont {E.}~\bibnamefont {Tonni}},\ }\bibfield  {title} {\bibinfo {title} {Modular {H}amiltonians for the massless {D}irac field in the presence of a defect},\ }\href {https://doi.org/10.1007/jhep03(2021)205} {\bibfield  {journal} {\bibinfo  {journal} {J. High Energy Phys.}\ }\textbf {\bibinfo {volume} {2021}}\bibinfo  {number} { (3)},\ \bibinfo {pages} {30}}\BibitemShut {NoStop}%
\bibitem [{\citenamefont {Rottoli}\ \emph {et~al.}(2023)\citenamefont {Rottoli}, \citenamefont {Murciano}, \citenamefont {Tonni},\ and\ \citenamefont {Calabrese}}]{RMTC-23}%
  \BibitemOpen
\bibfield  {number} {  }\bibfield  {author} {\bibinfo {author} {\bibfnamefont {F.}~\bibnamefont {Rottoli}}, \bibinfo {author} {\bibfnamefont {S.}~\bibnamefont {Murciano}}, \bibinfo {author} {\bibfnamefont {E.}~\bibnamefont {Tonni}},\ and\ \bibinfo {author} {\bibfnamefont {P.}~\bibnamefont {Calabrese}},\ }\bibfield  {title} {\bibinfo {title} {Entanglement and negativity {Hamiltonians} for the massless dirac field on the half line},\ }\href {https://doi.org/10.1088/1742-5468/acb262} {\bibfield  {journal} {\bibinfo  {journal} {J. Stat. Mech.}\ }\textbf {\bibinfo {volume} {2023}},\ \bibinfo {pages} {013103} (\bibinfo {year} {2023})}\BibitemShut {NoStop}%
\bibitem [{\citenamefont {Peschel}\ and\ \citenamefont {Eisler}(2009)}]{PE-09}%
  \BibitemOpen
  \bibfield  {author} {\bibinfo {author} {\bibfnamefont {I.}~\bibnamefont {Peschel}}\ and\ \bibinfo {author} {\bibfnamefont {V.}~\bibnamefont {Eisler}},\ }\bibfield  {title} {\bibinfo {title} {Reduced density matrices and entanglement entropy in free lattice models},\ }\href {https://doi.org/10.1088/1751-8113/42/50/504003} {\bibfield  {journal} {\bibinfo  {journal} {J. Phys. A: Math. Theor.}\ }\textbf {\bibinfo {volume} {42}},\ \bibinfo {pages} {504003} (\bibinfo {year} {2009})}\BibitemShut {NoStop}%
\bibitem [{\citenamefont {Eisler}\ and\ \citenamefont {Peschel}(2017)}]{EP-17}%
  \BibitemOpen
  \bibfield  {author} {\bibinfo {author} {\bibfnamefont {V.}~\bibnamefont {Eisler}}\ and\ \bibinfo {author} {\bibfnamefont {I.}~\bibnamefont {Peschel}},\ }\bibfield  {title} {\bibinfo {title} {Analytical results for the entanglement {Hamiltonian} of a free-fermion chain},\ }\href {https://doi.org/10.1088/1751-8121/aa76b5} {\bibfield  {journal} {\bibinfo  {journal} {J. Phys. A: Math. Theor.}\ }\textbf {\bibinfo {volume} {50}},\ \bibinfo {pages} {284003} (\bibinfo {year} {2017})}\BibitemShut {NoStop}%
\bibitem [{\citenamefont {Eisler}\ and\ \citenamefont {Peschel}(2018)}]{EP-18}%
  \BibitemOpen
  \bibfield  {author} {\bibinfo {author} {\bibfnamefont {V.}~\bibnamefont {Eisler}}\ and\ \bibinfo {author} {\bibfnamefont {I.}~\bibnamefont {Peschel}},\ }\bibfield  {title} {\bibinfo {title} {Properties of the entanglement {Hamiltonian} for finite free-fermion chains},\ }\href {https://doi.org/10.1088/1742-5468/aace2b} {\bibfield  {journal} {\bibinfo  {journal} {J. Stat. Mech.}\ }\textbf {\bibinfo {volume} {2018}},\ \bibinfo {pages} {104001} (\bibinfo {year} {2018})}\BibitemShut {NoStop}%
\bibitem [{\citenamefont {Eisler}\ \emph {et~al.}(2019)\citenamefont {Eisler}, \citenamefont {Tonni},\ and\ \citenamefont {Peschel}}]{ETP-19}%
  \BibitemOpen
  \bibfield  {author} {\bibinfo {author} {\bibfnamefont {V.}~\bibnamefont {Eisler}}, \bibinfo {author} {\bibfnamefont {E.}~\bibnamefont {Tonni}},\ and\ \bibinfo {author} {\bibfnamefont {I.}~\bibnamefont {Peschel}},\ }\bibfield  {title} {\bibinfo {title} {On the continuum limit of the entanglement {Hamiltonian}},\ }\href {https://doi.org/10.1088/1742-5468/ab1f0e} {\bibfield  {journal} {\bibinfo  {journal} {J. Stat. Mech.}\ }\textbf {\bibinfo {volume} {2019}},\ \bibinfo {pages} {073101} (\bibinfo {year} {2019})}\BibitemShut {NoStop}%
\bibitem [{\citenamefont {Eisler}\ \emph {et~al.}(2022)\citenamefont {Eisler}, \citenamefont {Tonni},\ and\ \citenamefont {Peschel}}]{ETP-22}%
  \BibitemOpen
  \bibfield  {author} {\bibinfo {author} {\bibfnamefont {V.}~\bibnamefont {Eisler}}, \bibinfo {author} {\bibfnamefont {E.}~\bibnamefont {Tonni}},\ and\ \bibinfo {author} {\bibfnamefont {I.}~\bibnamefont {Peschel}},\ }\bibfield  {title} {\bibinfo {title} {Local and non-local properties of the entanglement {Hamiltonian} for two disjoint intervals},\ }\href {https://doi.org/10.1088/1742-5468/ac8151} {\bibfield  {journal} {\bibinfo  {journal} {J. Stat. Mech.}\ }\textbf {\bibinfo {volume} {2022}},\ \bibinfo {pages} {083101} (\bibinfo {year} {2022})}\BibitemShut {NoStop}%
\bibitem [{\citenamefont {Rottoli}\ \emph {et~al.}(2022)\citenamefont {Rottoli}, \citenamefont {Scopa},\ and\ \citenamefont {Calabrese}}]{RSC-22}%
  \BibitemOpen
  \bibfield  {author} {\bibinfo {author} {\bibfnamefont {F.}~\bibnamefont {Rottoli}}, \bibinfo {author} {\bibfnamefont {S.}~\bibnamefont {Scopa}},\ and\ \bibinfo {author} {\bibfnamefont {P.}~\bibnamefont {Calabrese}},\ }\bibfield  {title} {\bibinfo {title} {Entanglement {Hamiltonian} during a domain wall melting in the free fermi chain},\ }\href {https://doi.org/10.1088/1742-5468/ac72a1} {\bibfield  {journal} {\bibinfo  {journal} {J. Stat. Mech.}\ }\textbf {\bibinfo {volume} {2022}},\ \bibinfo {pages} {063103} (\bibinfo {year} {2022})}\BibitemShut {NoStop}%
\bibitem [{\citenamefont {Bonsignori}\ and\ \citenamefont {Eisler}(2024)}]{BE-24}%
  \BibitemOpen
  \bibfield  {author} {\bibinfo {author} {\bibfnamefont {R.}~\bibnamefont {Bonsignori}}\ and\ \bibinfo {author} {\bibfnamefont {V.}~\bibnamefont {Eisler}},\ }\bibfield  {title} {\bibinfo {title} {Entanglement {Hamiltonian} for inhomogeneous free fermions},\ }\href {https://doi.org/10.1088/1751-8121/ad5501} {\bibfield  {journal} {\bibinfo  {journal} {J. Phys. A: Math. Theor.}\ }\textbf {\bibinfo {volume} {57}},\ \bibinfo {pages} {275001} (\bibinfo {year} {2024})}\BibitemShut {NoStop}%
\bibitem [{\citenamefont {Bernard}\ \emph {et~al.}(2024{\natexlab{a}})\citenamefont {Bernard}, \citenamefont {Bonsignori}, \citenamefont {Eisler}, \citenamefont {Parez},\ and\ \citenamefont {Vinet}}]{BBEPV-24}%
  \BibitemOpen
  \bibfield  {author} {\bibinfo {author} {\bibfnamefont {P.-A.}\ \bibnamefont {Bernard}}, \bibinfo {author} {\bibfnamefont {R.}~\bibnamefont {Bonsignori}}, \bibinfo {author} {\bibfnamefont {V.}~\bibnamefont {Eisler}}, \bibinfo {author} {\bibfnamefont {G.}~\bibnamefont {Parez}},\ and\ \bibinfo {author} {\bibfnamefont {L.}~\bibnamefont {Vinet}},\ }\href {https://arxiv.org/abs/2412.12021} {\bibinfo {title} {Entanglement {Hamiltonian} and orthogonal polynomials}} (\bibinfo {year} {2024}{\natexlab{a}}),\ \Eprint {https://arxiv.org/abs/2412.12021} {arXiv:2412.12021} \BibitemShut {NoStop}%
\bibitem [{\citenamefont {Giulio}\ and\ \citenamefont {Tonni}(2020)}]{DGT-20}%
  \BibitemOpen
  \bibfield  {author} {\bibinfo {author} {\bibfnamefont {G.~D.}\ \bibnamefont {Giulio}}\ and\ \bibinfo {author} {\bibfnamefont {E.}~\bibnamefont {Tonni}},\ }\bibfield  {title} {\bibinfo {title} {On entanglement {Hamiltonians} of an interval in massless harmonic chains},\ }\href {https://doi.org/10.1088/1742-5468/ab7129} {\bibfield  {journal} {\bibinfo  {journal} {J. Stat. Mech.}\ }\textbf {\bibinfo {volume} {2020}},\ \bibinfo {pages} {033102} (\bibinfo {year} {2020})}\BibitemShut {NoStop}%
\bibitem [{\citenamefont {Javerzat}\ and\ \citenamefont {Tonni}(2022)}]{JT-22}%
  \BibitemOpen
  \bibfield  {author} {\bibinfo {author} {\bibfnamefont {N.}~\bibnamefont {Javerzat}}\ and\ \bibinfo {author} {\bibfnamefont {E.}~\bibnamefont {Tonni}},\ }\bibfield  {title} {\bibinfo {title} {On the continuum limit of the entanglement {Hamiltonian} of a sphere for the free massless scalar field},\ }\href {https://doi.org/10.1007/JHEP02(2022)086} {\bibfield  {journal} {\bibinfo  {journal} {J. High Energy Phys.}\ }\textbf {\bibinfo {volume} {2022}}}\BibitemShut {NoStop}%
\bibitem [{\citenamefont {Gentile}\ \emph {et~al.}(2025)\citenamefont {Gentile}, \citenamefont {Rotaru},\ and\ \citenamefont {Tonni}}]{GRT-25}%
  \BibitemOpen
  \bibfield  {author} {\bibinfo {author} {\bibfnamefont {F.}~\bibnamefont {Gentile}}, \bibinfo {author} {\bibfnamefont {A.}~\bibnamefont {Rotaru}},\ and\ \bibinfo {author} {\bibfnamefont {E.}~\bibnamefont {Tonni}},\ }\bibfield  {title} {\bibinfo {title} {Entanglement {Hamiltonian} of two disjoint blocks in the harmonic chain},\ }\href {https://doi.org/10.1088/1742-5468/addaa7} {\bibfield  {journal} {\bibinfo  {journal} {J. Stat. Mech.}\ }\textbf {\bibinfo {volume} {2025}},\ \bibinfo {pages} {073102} (\bibinfo {year} {2025})}\BibitemShut {NoStop}%
\bibitem [{\citenamefont {Calabrese}\ and\ \citenamefont {Cardy}(2005)}]{CC-05}%
  \BibitemOpen
  \bibfield  {author} {\bibinfo {author} {\bibfnamefont {P.}~\bibnamefont {Calabrese}}\ and\ \bibinfo {author} {\bibfnamefont {J.}~\bibnamefont {Cardy}},\ }\bibfield  {title} {\bibinfo {title} {Evolution of entanglement entropy in one-dimensional systems},\ }\href {https://doi.org/10.1088/1742-5468/2005/04/P04010} {\bibfield  {journal} {\bibinfo  {journal} {J. Stat. Mech.}\ }\textbf {\bibinfo {volume} {2005}},\ \bibinfo {pages} {P04010} (\bibinfo {year} {2005})}\BibitemShut {NoStop}%
\bibitem [{\citenamefont {Di~Giulio}\ \emph {et~al.}(2019)\citenamefont {Di~Giulio}, \citenamefont {Arias},\ and\ \citenamefont {Tonni}}]{dGAT-19}%
  \BibitemOpen
  \bibfield  {author} {\bibinfo {author} {\bibfnamefont {G.}~\bibnamefont {Di~Giulio}}, \bibinfo {author} {\bibfnamefont {R.}~\bibnamefont {Arias}},\ and\ \bibinfo {author} {\bibfnamefont {E.}~\bibnamefont {Tonni}},\ }\bibfield  {title} {\bibinfo {title} {Entanglement {Hamiltonians} in 1d free lattice models after a global quantum quench},\ }\href {https://doi.org/10.1088/1742-5468/ab4e8f} {\bibfield  {journal} {\bibinfo  {journal} {J. Stat. Mech.}\ }\textbf {\bibinfo {volume} {2019}},\ \bibinfo {pages} {123103} (\bibinfo {year} {2019})}\BibitemShut {NoStop}%
\bibitem [{\citenamefont {Rottoli}\ \emph {et~al.}(2025)\citenamefont {Rottoli}, \citenamefont {Rylands},\ and\ \citenamefont {Calabrese}}]{RRC-24}%
  \BibitemOpen
  \bibfield  {author} {\bibinfo {author} {\bibfnamefont {F.}~\bibnamefont {Rottoli}}, \bibinfo {author} {\bibfnamefont {C.}~\bibnamefont {Rylands}},\ and\ \bibinfo {author} {\bibfnamefont {P.}~\bibnamefont {Calabrese}},\ }\bibfield  {title} {\bibinfo {title} {Entanglement {Hamiltonians} and the quasiparticle picture},\ }\href {https://doi.org/10.1103/PhysRevB.111.L140302} {\bibfield  {journal} {\bibinfo  {journal} {Phys. Rev. B}\ }\textbf {\bibinfo {volume} {111}},\ \bibinfo {pages} {L140302} (\bibinfo {year} {2025})}\BibitemShut {NoStop}%
\bibitem [{\citenamefont {Travaglino}\ \emph {et~al.}(2025)\citenamefont {Travaglino}, \citenamefont {Rylands},\ and\ \citenamefont {Calabrese}}]{TRC-25}%
  \BibitemOpen
  \bibfield  {author} {\bibinfo {author} {\bibfnamefont {R.}~\bibnamefont {Travaglino}}, \bibinfo {author} {\bibfnamefont {C.}~\bibnamefont {Rylands}},\ and\ \bibinfo {author} {\bibfnamefont {P.}~\bibnamefont {Calabrese}},\ }\bibfield  {title} {\bibinfo {title} {Quasiparticle picture for entanglement {Hamiltonians} in higher dimensions},\ }\href {https://doi.org/10.1088/1742-5468/adb7d3} {\bibfield  {journal} {\bibinfo  {journal} {J. Stat. Mech.}\ }\textbf {\bibinfo {volume} {2025}},\ \bibinfo {pages} {033102} (\bibinfo {year} {2025})}\BibitemShut {NoStop}%
\bibitem [{\citenamefont {Eisler}\ and\ \citenamefont {Peschel}(2007)}]{EP-07}%
  \BibitemOpen
  \bibfield  {author} {\bibinfo {author} {\bibfnamefont {V.}~\bibnamefont {Eisler}}\ and\ \bibinfo {author} {\bibfnamefont {I.}~\bibnamefont {Peschel}},\ }\bibfield  {title} {\bibinfo {title} {Evolution of entanglement after a local quench},\ }\href {https://doi.org/10.1088/1742-5468/2007/06/P06005} {\bibfield  {journal} {\bibinfo  {journal} {J. Stat. Mech.}\ }\textbf {\bibinfo {volume} {2007}},\ \bibinfo {pages} {P06005} (\bibinfo {year} {2007})}\BibitemShut {NoStop}%
\bibitem [{\citenamefont {Eisler}\ \emph {et~al.}(2008)\citenamefont {Eisler}, \citenamefont {Karevski}, \citenamefont {Platini},\ and\ \citenamefont {Peschel}}]{EKPP-08}%
  \BibitemOpen
  \bibfield  {author} {\bibinfo {author} {\bibfnamefont {V.}~\bibnamefont {Eisler}}, \bibinfo {author} {\bibfnamefont {D.}~\bibnamefont {Karevski}}, \bibinfo {author} {\bibfnamefont {T.}~\bibnamefont {Platini}},\ and\ \bibinfo {author} {\bibfnamefont {I.}~\bibnamefont {Peschel}},\ }\bibfield  {title} {\bibinfo {title} {Entanglement evolution after connecting finite to infinite quantum chains},\ }\href {https://doi.org/10.1088/1742-5468/2008/01/P01023} {\bibfield  {journal} {\bibinfo  {journal} {J. Stat. Mech.}\ }\textbf {\bibinfo {volume} {2008}},\ \bibinfo {pages} {P01023} (\bibinfo {year} {2008})}\BibitemShut {NoStop}%
\bibitem [{\citenamefont {Igl\'oi}\ \emph {et~al.}(2009)\citenamefont {Igl\'oi}, \citenamefont {Szatm\'ari},\ and\ \citenamefont {Lin}}]{ISL-09}%
  \BibitemOpen
  \bibfield  {author} {\bibinfo {author} {\bibfnamefont {F.}~\bibnamefont {Igl\'oi}}, \bibinfo {author} {\bibfnamefont {Z.}~\bibnamefont {Szatm\'ari}},\ and\ \bibinfo {author} {\bibfnamefont {Y.-C.}\ \bibnamefont {Lin}},\ }\bibfield  {title} {\bibinfo {title} {Entanglement entropy with localized and extended interface defects},\ }\href {https://doi.org/10.1103/PhysRevB.80.024405} {\bibfield  {journal} {\bibinfo  {journal} {Phys. Rev. B}\ }\textbf {\bibinfo {volume} {80}},\ \bibinfo {pages} {024405} (\bibinfo {year} {2009})}\BibitemShut {NoStop}%
\bibitem [{\citenamefont {Eisler}\ and\ \citenamefont {Peschel}(2012)}]{EP-12}%
  \BibitemOpen
  \bibfield  {author} {\bibinfo {author} {\bibfnamefont {V.}~\bibnamefont {Eisler}}\ and\ \bibinfo {author} {\bibfnamefont {I.}~\bibnamefont {Peschel}},\ }\bibfield  {title} {\bibinfo {title} {On entanglement evolution across defects in critical chains},\ }\href {https://doi.org/10.1209/0295-5075/99/20001} {\bibfield  {journal} {\bibinfo  {journal} {Europhys. Lett.}\ }\textbf {\bibinfo {volume} {99}},\ \bibinfo {pages} {20001} (\bibinfo {year} {2012})}\BibitemShut {NoStop}%
\bibitem [{\citenamefont {Collura}\ and\ \citenamefont {Calabrese}(2013)}]{CC-13}%
  \BibitemOpen
  \bibfield  {author} {\bibinfo {author} {\bibfnamefont {M.}~\bibnamefont {Collura}}\ and\ \bibinfo {author} {\bibfnamefont {P.}~\bibnamefont {Calabrese}},\ }\bibfield  {title} {\bibinfo {title} {Entanglement evolution across defects in critical anisotropic {Heisenberg} chains},\ }\href {https://doi.org/10.1088/1751-8113/46/17/175001} {\bibfield  {journal} {\bibinfo  {journal} {J. Phys. A: Math. Theor.}\ }\textbf {\bibinfo {volume} {46}},\ \bibinfo {pages} {175001} (\bibinfo {year} {2013})}\BibitemShut {NoStop}%
\bibitem [{\citenamefont {Vasseur}\ and\ \citenamefont {Saleur}(2017)}]{VS-17}%
  \BibitemOpen
  \bibfield  {author} {\bibinfo {author} {\bibfnamefont {R.}~\bibnamefont {Vasseur}}\ and\ \bibinfo {author} {\bibfnamefont {H.}~\bibnamefont {Saleur}},\ }\bibfield  {title} {\bibinfo {title} {{Universal Entanglement Dynamics following a Local Quench}},\ }\href {https://doi.org/10.21468/SciPostPhys.3.1.001} {\bibfield  {journal} {\bibinfo  {journal} {SciPost Phys.}\ }\textbf {\bibinfo {volume} {3}},\ \bibinfo {pages} {001} (\bibinfo {year} {2017})}\BibitemShut {NoStop}%
\bibitem [{\citenamefont {Di~Giulio}\ and\ \citenamefont {Tonni}(2021)}]{GT-21}%
  \BibitemOpen
  \bibfield  {author} {\bibinfo {author} {\bibfnamefont {G.}~\bibnamefont {Di~Giulio}}\ and\ \bibinfo {author} {\bibfnamefont {E.}~\bibnamefont {Tonni}},\ }\bibfield  {title} {\bibinfo {title} {Subsystem complexity after a local quantum quench},\ }\href {https://doi.org/10.1007/JHEP08(2021)135} {\bibfield  {journal} {\bibinfo  {journal} {J. High Energy Phys.}\ }\textbf {\bibinfo {volume} {2021}}}\BibitemShut {NoStop}%
\bibitem [{\citenamefont {Calabrese}\ and\ \citenamefont {Cardy}(2007)}]{CC-07}%
  \BibitemOpen
  \bibfield  {author} {\bibinfo {author} {\bibfnamefont {P.}~\bibnamefont {Calabrese}}\ and\ \bibinfo {author} {\bibfnamefont {J.}~\bibnamefont {Cardy}},\ }\bibfield  {title} {\bibinfo {title} {Entanglement and correlation functions following a local quench: a conformal field theory approach},\ }\href {https://doi.org/10.1088/1742-5468/2007/10/P10004} {\bibfield  {journal} {\bibinfo  {journal} {J. Stat. Mech.}\ }\textbf {\bibinfo {volume} {2007}},\ \bibinfo {pages} {P10004} (\bibinfo {year} {2007})}\BibitemShut {NoStop}%
\bibitem [{\citenamefont {Stéphan}\ and\ \citenamefont {Dubail}(2011)}]{SD-11}%
  \BibitemOpen
  \bibfield  {author} {\bibinfo {author} {\bibfnamefont {J.-M.}\ \bibnamefont {Stéphan}}\ and\ \bibinfo {author} {\bibfnamefont {J.}~\bibnamefont {Dubail}},\ }\bibfield  {title} {\bibinfo {title} {Local quantum quenches in critical one-dimensional systems: entanglement, the {Loschmidt} echo, and light-cone effects},\ }\href {https://doi.org/10.1088/1742-5468/2011/08/P08019} {\bibfield  {journal} {\bibinfo  {journal} {J. Stat. Mech.}\ }\textbf {\bibinfo {volume} {2011}},\ \bibinfo {pages} {P08019} (\bibinfo {year} {2011})}\BibitemShut {NoStop}%
\bibitem [{\citenamefont {Cardy}(2011)}]{Cardy-11}%
  \BibitemOpen
  \bibfield  {author} {\bibinfo {author} {\bibfnamefont {J.}~\bibnamefont {Cardy}},\ }\bibfield  {title} {\bibinfo {title} {Measuring entanglement using quantum quenches},\ }\href {https://doi.org/10.1103/PhysRevLett.106.150404} {\bibfield  {journal} {\bibinfo  {journal} {Phys. Rev. Lett.}\ }\textbf {\bibinfo {volume} {106}},\ \bibinfo {pages} {150404} (\bibinfo {year} {2011})}\BibitemShut {NoStop}%
\bibitem [{\citenamefont {Asplund}\ and\ \citenamefont {Bernamonti}(2014)}]{AB-14}%
  \BibitemOpen
  \bibfield  {author} {\bibinfo {author} {\bibfnamefont {C.~T.}\ \bibnamefont {Asplund}}\ and\ \bibinfo {author} {\bibfnamefont {A.}~\bibnamefont {Bernamonti}},\ }\bibfield  {title} {\bibinfo {title} {Mutual information after a local quench in conformal field theory},\ }\href {https://doi.org/10.1103/PhysRevD.89.066015} {\bibfield  {journal} {\bibinfo  {journal} {Phys. Rev. D}\ }\textbf {\bibinfo {volume} {89}},\ \bibinfo {pages} {066015} (\bibinfo {year} {2014})}\BibitemShut {NoStop}%
\bibitem [{\citenamefont {Wen}\ \emph {et~al.}(2015)\citenamefont {Wen}, \citenamefont {Chang},\ and\ \citenamefont {Ryu}}]{WCR-15}%
  \BibitemOpen
  \bibfield  {author} {\bibinfo {author} {\bibfnamefont {X.}~\bibnamefont {Wen}}, \bibinfo {author} {\bibfnamefont {P.-Y.}\ \bibnamefont {Chang}},\ and\ \bibinfo {author} {\bibfnamefont {S.}~\bibnamefont {Ryu}},\ }\bibfield  {title} {\bibinfo {title} {Entanglement negativity after a local quantum quench in conformal field theories},\ }\href {https://doi.org/10.1103/PhysRevB.92.075109} {\bibfield  {journal} {\bibinfo  {journal} {Phys. Rev. B}\ }\textbf {\bibinfo {volume} {92}},\ \bibinfo {pages} {075109} (\bibinfo {year} {2015})}\BibitemShut {NoStop}%
\bibitem [{\citenamefont {Caputa}\ and\ \citenamefont {Giulio}(2025)}]{CdG-25}%
  \BibitemOpen
  \bibfield  {author} {\bibinfo {author} {\bibfnamefont {P.}~\bibnamefont {Caputa}}\ and\ \bibinfo {author} {\bibfnamefont {G.~D.}\ \bibnamefont {Giulio}},\ }\href {https://arxiv.org/abs/2502.19485} {\bibinfo {title} {Local quenches from a krylov perspective}} (\bibinfo {year} {2025}),\ \Eprint {https://arxiv.org/abs/2502.19485} {arXiv:2502.19485} \BibitemShut {NoStop}%
\bibitem [{\citenamefont {Nozaki}\ \emph {et~al.}(2013)\citenamefont {Nozaki}, \citenamefont {Numasawa},\ and\ \citenamefont {Takayanagi}}]{NNT-13}%
  \BibitemOpen
  \bibfield  {author} {\bibinfo {author} {\bibfnamefont {M.}~\bibnamefont {Nozaki}}, \bibinfo {author} {\bibfnamefont {T.}~\bibnamefont {Numasawa}},\ and\ \bibinfo {author} {\bibfnamefont {T.}~\bibnamefont {Takayanagi}},\ }\bibfield  {title} {\bibinfo {title} {Holographic local quenches and entanglement density},\ }\href {https://doi.org/10.1007/JHEP05(2013)080} {\bibfield  {journal} {\bibinfo  {journal} {J. High Energy Phys.}\ }\textbf {\bibinfo {volume} {2013}},\ \bibinfo {pages} {80}}\BibitemShut {NoStop}%
\bibitem [{\citenamefont {Ugajin}(2013)}]{U-13}%
  \BibitemOpen
  \bibfield  {author} {\bibinfo {author} {\bibfnamefont {T.}~\bibnamefont {Ugajin}},\ }\href {https://arxiv.org/abs/1311.2562} {\bibinfo {title} {Two dimensional quantum quenches and holography}} (\bibinfo {year} {2013}),\ \Eprint {https://arxiv.org/abs/1311.2562} {arXiv:1311.2562} \BibitemShut {NoStop}%
\bibitem [{\citenamefont {Shimaji}\ \emph {et~al.}(2019)\citenamefont {Shimaji}, \citenamefont {Takayanagi},\ and\ \citenamefont {Wei}}]{STW-19}%
  \BibitemOpen
  \bibfield  {author} {\bibinfo {author} {\bibfnamefont {T.}~\bibnamefont {Shimaji}}, \bibinfo {author} {\bibfnamefont {T.}~\bibnamefont {Takayanagi}},\ and\ \bibinfo {author} {\bibfnamefont {Z.}~\bibnamefont {Wei}},\ }\bibfield  {title} {\bibinfo {title} {Holographic quantum circuits from splitting/joining local quenches},\ }\href {https://doi.org/10.1007/JHEP03(2019)165} {\bibfield  {journal} {\bibinfo  {journal} {J. High Energy Phys.}\ }\textbf {\bibinfo {volume} {2019}},\ \bibinfo {pages} {165}}\BibitemShut {NoStop}%
\bibitem [{\citenamefont {Kudler-Flam}\ \emph {et~al.}(2024)\citenamefont {Kudler-Flam}, \citenamefont {Nozaki}, \citenamefont {Numasawa}, \citenamefont {Ryu},\ and\ \citenamefont {Tan}}]{KFNNRT-23}%
  \BibitemOpen
  \bibfield  {author} {\bibinfo {author} {\bibfnamefont {J.}~\bibnamefont {Kudler-Flam}}, \bibinfo {author} {\bibfnamefont {M.}~\bibnamefont {Nozaki}}, \bibinfo {author} {\bibfnamefont {T.}~\bibnamefont {Numasawa}}, \bibinfo {author} {\bibfnamefont {S.}~\bibnamefont {Ryu}},\ and\ \bibinfo {author} {\bibfnamefont {M.~T.}\ \bibnamefont {Tan}},\ }\bibfield  {title} {\bibinfo {title} {Bridging two quantum quench problems -- local joining quantum quench and \text{M}\"obius quench -- and their holographic dual descriptions},\ }\href {https://doi.org/10.1007/JHEP08(2024)213} {\bibfield  {journal} {\bibinfo  {journal} {J. High Energy Phys.}\ }\textbf {\bibinfo {volume} {2024}}}\BibitemShut {NoStop}%
\bibitem [{\citenamefont {Fagotti}\ and\ \citenamefont {Calabrese}(2011)}]{FC-11}%
  \BibitemOpen
  \bibfield  {author} {\bibinfo {author} {\bibfnamefont {M.}~\bibnamefont {Fagotti}}\ and\ \bibinfo {author} {\bibfnamefont {P.}~\bibnamefont {Calabrese}},\ }\bibfield  {title} {\bibinfo {title} {Universal parity effects in the entanglement entropy of {XX} chains with open boundary conditions},\ }\href {https://doi.org/10.1088/1742-5468/2011/01/P01017} {\bibfield  {journal} {\bibinfo  {journal} {J. Stat. Mech.}\ }\textbf {\bibinfo {volume} {2011}},\ \bibinfo {pages} {P01017} (\bibinfo {year} {2011})}\BibitemShut {NoStop}%
\bibitem [{\citenamefont {Peschel}(2003)}]{Peschel03}%
  \BibitemOpen
  \bibfield  {author} {\bibinfo {author} {\bibfnamefont {I.}~\bibnamefont {Peschel}},\ }\bibfield  {title} {\bibinfo {title} {Calculation of reduced density matrices from correlation functions},\ }\href {https://doi.org/10.1088/0305-4470/36/14/101} {\bibfield  {journal} {\bibinfo  {journal} {J. Phys. A: Math. Theor.}\ }\textbf {\bibinfo {volume} {36}},\ \bibinfo {pages} {L205} (\bibinfo {year} {2003})}\BibitemShut {NoStop}%
\bibitem [{\citenamefont {Peschel}(2004)}]{P-04}%
  \BibitemOpen
  \bibfield  {author} {\bibinfo {author} {\bibfnamefont {I.}~\bibnamefont {Peschel}},\ }\bibfield  {title} {\bibinfo {title} {On the reduced density matrix for a chain of free electrons},\ }\href {https://doi.org/10.1088/1742-5468/2004/06/P06004} {\bibfield  {journal} {\bibinfo  {journal} {J. Stat. Mech.}\ }\textbf {\bibinfo {volume} {2004}},\ \bibinfo {pages} {P06004} (\bibinfo {year} {2004})}\BibitemShut {NoStop}%
\bibitem [{\citenamefont {Crampé}\ \emph {et~al.}(2019)\citenamefont {Crampé}, \citenamefont {Nepomechie},\ and\ \citenamefont {Vinet}}]{CNV-19}%
  \BibitemOpen
  \bibfield  {author} {\bibinfo {author} {\bibfnamefont {N.}~\bibnamefont {Crampé}}, \bibinfo {author} {\bibfnamefont {R.~I.}\ \bibnamefont {Nepomechie}},\ and\ \bibinfo {author} {\bibfnamefont {L.}~\bibnamefont {Vinet}},\ }\bibfield  {title} {\bibinfo {title} {Free-fermion entanglement and orthogonal polynomials},\ }\href {https://doi.org/10.1088/1742-5468/ab3787} {\bibfield  {journal} {\bibinfo  {journal} {J. Stat. Mech.}\ }\textbf {\bibinfo {volume} {2019}},\ \bibinfo {pages} {093101} (\bibinfo {year} {2019})}\BibitemShut {NoStop}%
\bibitem [{\citenamefont {Bernard}\ \emph {et~al.}(2024{\natexlab{b}})\citenamefont {Bernard}, \citenamefont {Crampé}, \citenamefont {Nepomechie}, \citenamefont {Parez},\ and\ \citenamefont {Vinet}}]{BCNPV-24}%
  \BibitemOpen
  \bibfield  {author} {\bibinfo {author} {\bibfnamefont {P.-A.}\ \bibnamefont {Bernard}}, \bibinfo {author} {\bibfnamefont {N.}~\bibnamefont {Crampé}}, \bibinfo {author} {\bibfnamefont {R.~I.}\ \bibnamefont {Nepomechie}}, \bibinfo {author} {\bibfnamefont {G.}~\bibnamefont {Parez}},\ and\ \bibinfo {author} {\bibfnamefont {L.}~\bibnamefont {Vinet}},\ }\href {https://arxiv.org/abs/2401.07150} {\bibinfo {title} {Entanglement of free-fermion systems, signal processing and algebraic combinatorics}} (\bibinfo {year} {2024}{\natexlab{b}}),\ \Eprint {https://arxiv.org/abs/2401.07150} {arXiv:2401.07150} \BibitemShut {NoStop}%
\bibitem [{\citenamefont {Eisler}(2025)}]{Eis-25}%
  \BibitemOpen
  \bibfield  {author} {\bibinfo {author} {\bibfnamefont {V.}~\bibnamefont {Eisler}},\ }\bibfield  {title} {\bibinfo {title} {On the {Bisognano–Wichmann} entanglement {Hamiltonian} of nonrelativistic fermions},\ }\href {https://doi.org/10.1088/1742-5468/ad9c4f} {\bibfield  {journal} {\bibinfo  {journal} {J. Stat. Mech.}\ }\textbf {\bibinfo {volume} {2025}},\ \bibinfo {pages} {013101} (\bibinfo {year} {2025})}\BibitemShut {NoStop}%
\bibitem [{\citenamefont {Dalmonte}\ \emph {et~al.}(2018)\citenamefont {Dalmonte}, \citenamefont {Vermersch},\ and\ \citenamefont {Zoller}}]{DVZ-18}%
  \BibitemOpen
  \bibfield  {author} {\bibinfo {author} {\bibfnamefont {M.}~\bibnamefont {Dalmonte}}, \bibinfo {author} {\bibfnamefont {B.}~\bibnamefont {Vermersch}},\ and\ \bibinfo {author} {\bibfnamefont {P.}~\bibnamefont {Zoller}},\ }\bibfield  {title} {\bibinfo {title} {Quantum simulation and spectroscopy of entanglement {Hamiltonians}},\ }\href {https://doi.org/10.1038/s41567-018-0151-7} {\bibfield  {journal} {\bibinfo  {journal} {Nat. Phys.}\ }\textbf {\bibinfo {volume} {14}},\ \bibinfo {pages} {827} (\bibinfo {year} {2018})}\BibitemShut {NoStop}%
\bibitem [{\citenamefont {Mendes-Santos}\ \emph {et~al.}(2019)\citenamefont {Mendes-Santos}, \citenamefont {Giudici}, \citenamefont {Dalmonte},\ and\ \citenamefont {Rajabpour}}]{GMSCD-18}%
  \BibitemOpen
  \bibfield  {author} {\bibinfo {author} {\bibfnamefont {T.}~\bibnamefont {Mendes-Santos}}, \bibinfo {author} {\bibfnamefont {G.}~\bibnamefont {Giudici}}, \bibinfo {author} {\bibfnamefont {M.}~\bibnamefont {Dalmonte}},\ and\ \bibinfo {author} {\bibfnamefont {M.~A.}\ \bibnamefont {Rajabpour}},\ }\bibfield  {title} {\bibinfo {title} {Entanglement {Hamiltonian} of quantum critical chains and conformal field theories},\ }\href {https://doi.org/10.1103/PhysRevB.100.155122} {\bibfield  {journal} {\bibinfo  {journal} {Phys. Rev. B}\ }\textbf {\bibinfo {volume} {100}},\ \bibinfo {pages} {155122} (\bibinfo {year} {2019})}\BibitemShut {NoStop}%
\bibitem [{\citenamefont {Kokail}\ \emph {et~al.}(2021)\citenamefont {Kokail}, \citenamefont {van Bijnen}, \citenamefont {Elben}, \citenamefont {Vermersch},\ and\ \citenamefont {Zoller}}]{Kokailetal21a}%
  \BibitemOpen
  \bibfield  {author} {\bibinfo {author} {\bibfnamefont {C.}~\bibnamefont {Kokail}}, \bibinfo {author} {\bibfnamefont {R.}~\bibnamefont {van Bijnen}}, \bibinfo {author} {\bibfnamefont {A.}~\bibnamefont {Elben}}, \bibinfo {author} {\bibfnamefont {B.}~\bibnamefont {Vermersch}},\ and\ \bibinfo {author} {\bibfnamefont {P.}~\bibnamefont {Zoller}},\ }\bibfield  {title} {\bibinfo {title} {Entanglement {Hamiltonian} tomography in quantum simulation},\ }\href {https://doi.org/10.1038/s41567-021-01260-w} {\bibfield  {journal} {\bibinfo  {journal} {Nat. Phys.}\ }\textbf {\bibinfo {volume} {17}},\ \bibinfo {pages} {936} (\bibinfo {year} {2021})}\BibitemShut {NoStop}%
\bibitem [{\citenamefont {Joshi}\ \emph {et~al.}(2023)\citenamefont {Joshi}, \citenamefont {Kokail}, \citenamefont {van Bijnen}, \citenamefont {Kranzl}, \citenamefont {Zache}, \citenamefont {Blatt}, \citenamefont {Roos},\ and\ \citenamefont {Zoller}}]{Joshietal23}%
  \BibitemOpen
  \bibfield  {author} {\bibinfo {author} {\bibfnamefont {M.~K.}\ \bibnamefont {Joshi}}, \bibinfo {author} {\bibfnamefont {C.}~\bibnamefont {Kokail}}, \bibinfo {author} {\bibfnamefont {R.}~\bibnamefont {van Bijnen}}, \bibinfo {author} {\bibfnamefont {F.}~\bibnamefont {Kranzl}}, \bibinfo {author} {\bibfnamefont {T.~V.}\ \bibnamefont {Zache}}, \bibinfo {author} {\bibfnamefont {R.}~\bibnamefont {Blatt}}, \bibinfo {author} {\bibfnamefont {C.~F.}\ \bibnamefont {Roos}},\ and\ \bibinfo {author} {\bibfnamefont {P.}~\bibnamefont {Zoller}},\ }\bibfield  {title} {\bibinfo {title} {Exploring large-scale entanglement in quantum simulation},\ }\href {https://doi.org/10.1038/s41586-023-06768-0} {\bibfield  {journal} {\bibinfo  {journal} {Nature}\ }\textbf {\bibinfo {volume} {624}},\ \bibinfo {pages} {539} (\bibinfo {year} {2023})}\BibitemShut {NoStop}%
\end{thebibliography}
\end{document}